\documentclass[preprint2]{aastex}

\usepackage[caption=false]{subfig}
\usepackage{amsmath}

\newcommand{\myemail}{sspolaor@nrao.edu}
\newcommand{\fgearth}{f}

\newcommand{\fgrest}{f_{\rm r}}
\newcommand{\fbrest}{f_{\rm br}}
\newcommand{\eg}{e.\,g.}
\newcommand{\tobs}{T}
\newcommand{\yHz}{{\rm year^{-1}}}
\newcommand{\ayr}{A_{\rm 1yr}}
\newcommand{\mc}{M_{\rm c}}
\newcommand{\hc}{h_{\rm c}}

\newcommand{\msun}{\rm M_\odot}
\newcommand{\sn}{\sigma_{\rm n}}

\slugcomment{Submitted to PASP}

\shorttitle{Pulsar Timing Arrays}
\shortauthors{Burke-Spolaor}

\begin{document}

%% LaTeX will automatically break titles if they run longer than
%% one line. However, you may use \\ to force a line break if
%% you desire.

\title{Gravitational-Wave Detection and Astrophysics with Pulsar Timing Arrays}
%: Status and Prospects}

%% Use \author, \affil, and the \and command to format
%% author and affiliation information.
%% Note that \email has replaced the old \authoremail command
%% from AASTeX v4.0. You can use \email to mark an email address
%% anywhere in the paper, not just in the front matter.
%% As in the title, use \\ to force line breaks.

\author{S. Burke-Spolaor\altaffilmark{1}}
\affil{National Radio Astronomy Observatory, PO Box O, 1003 Lopezville Rd, Socorro, NM 87801-0387, USA}
\email{\myemail}
\altaffiltext{1}{Jansky Fellow}

\begin{abstract}
We have begun an exciting era for gravitational wave detection, as several world-leading experiments are breaching the threshold of anticipated signal strengths. Pulsar timing arrays (PTAs) are pan-Galactic gravitational wave detectors that are already cutting into the expected strength of gravitational waves from cosmic strings and binary supermassive black holes in the nHz-$\mu$Hz gravitational wave band. These limits are leading to constraints on the evolutionary state of the Universe. Here, we provide a broad review of this field, from how pulsars are used as tools for detection, to astrophysical sources of uncertainty in the signals PTAs aim to see, to the primary current challenge areas for PTA work. 
This review aims to provide an up-to-date reference point for new parties interested in the field of gravitational wave detection via pulsar timing.
\end{abstract}

%% Keywords should appear after the \end{abstract} command. The uncommented
%% example has been keyed in ApJ style. See the instructions to authors
%% for the journal to which you are submitting your paper to determine
%% what keyword punctuation is appropriate.


\section{Introduction}

%Potential co-authors?:
%George Hobbs
%Xavi Siemens
%Joe Simon
%\emph{General intro as to the purpose of the review and mathematical preliminaries: PTA detection of GWs}

The discovery of pulsars in 1968 initiated the flourishing field of pulsar astronomy, opening up the use of pulsars as tools to perform a vast range of experiments. The beams of light emitted from pulsars' magnetic poles mark their rotational phase at each turn. Coupled with their exceptional rotational stability, most pulsars are massive, moving, precisely ticking test masses: an ideal relativity tester.

\begin{figure*}
\centering
\vspace{-4mm}
\includegraphics[trim=0cm 0cm 0mm 0cm, clip, width=0.97\textwidth]{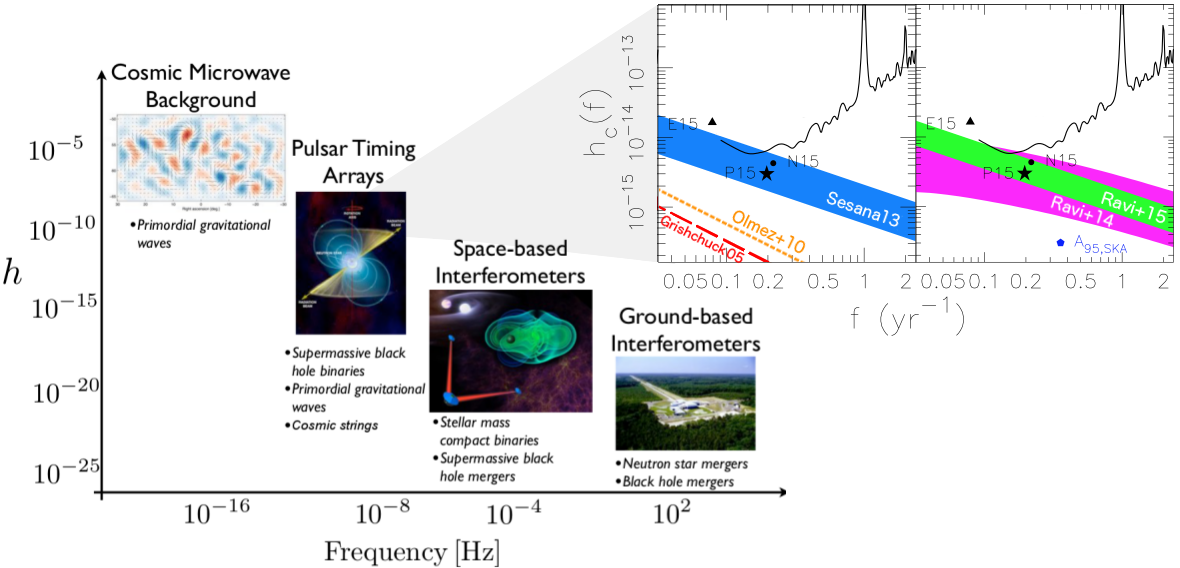}
\vspace{-3mm}
\caption{A conceptual strain spectrum, showing the complementarity of GW detection techniques adapted from a figure by NANOGrav. The anticipated sources, listed for each frequency range, are also complementary. The inset figure shows an enhanced view of the predicted GWB and limits in the PTA band (adapted and updated from \citealt{pptascience2}). The 68\% uncertainty range of several BSMBH simulations are indicated (filled curves), as are representative spectra from inflationary GWs (red long-dashed) and cosmic strings (orange short-dashed). The best limits are also shown; points are from PPTA, NANOGrav, and the EPTA \citep[][respectively]{pptascience2,9yrNANOlimit,lentati+15}; the curve is from PPTA \citep{pptascience2}.}
\vspace{-3mm}
\label{fig:overview}
\label{fig:turnovers}
\end{figure*}

One of the leading uses of pulsars is in ``Pulsar Timing Arrays'' (PTAs), which aim to detect gravitational radiation via a pan-Galactic detector.
The basic premise of the PTA is to detect variations in the arrival phases of a pulsar's tick, and search for correlations in these variations in a set of pulsars distributed across the sky. This method of gravitational wave (GW) detection is both competitive and complementary to other experiments pursuing the first direct detection of GWs (Fig.\,\ref{fig:overview}).

This review seeks to provide an accessible broad-strokes outline of the background, current state, and major efforts of gravitational wave detection with PTAs, for anyone new to, or wishing to enter, this field.

%This review aims to provide an up-to-date reference point for new parties interested in the field of gravitational wave detection with pulsar timing. 
%We also wish to include some standardly shown figures which need to be published.... for instance moore's law for RMS residuals?
% Mangum's words: ``aim to provide a ``go-to'' reference point for those wanting to update their understanding of a particular topic or for those wanting to understand and enter the field for the first time (i.e. students).''

%\section{Pulsars and Gravity}
\section{Basics}
\subsection{Timing data and GW frequency range}\label{sec:basics}
\begin{figure*}
\centering
\includegraphics[width=0.45\textwidth]{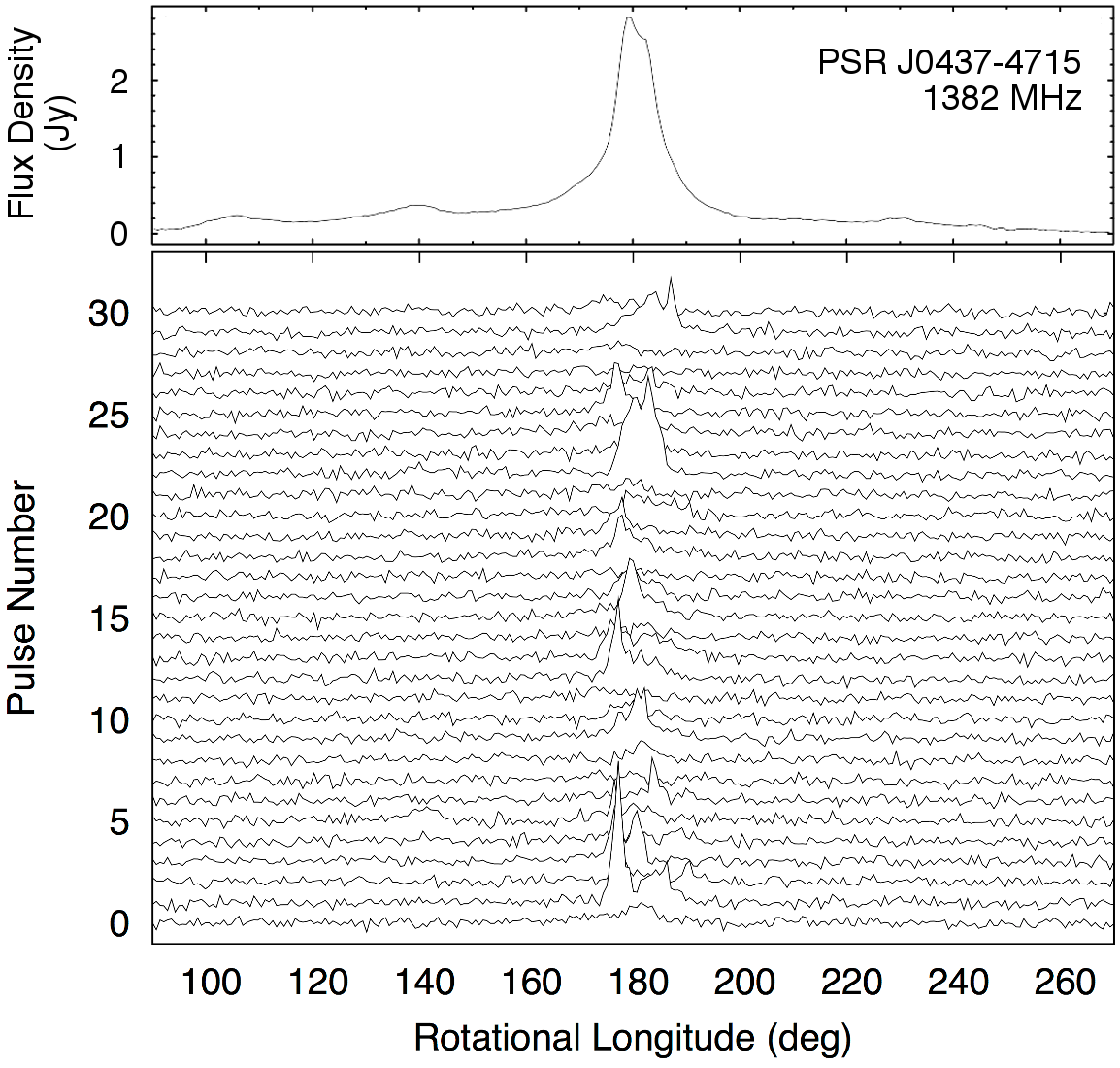}~~
\includegraphics[width=0.45\textwidth]{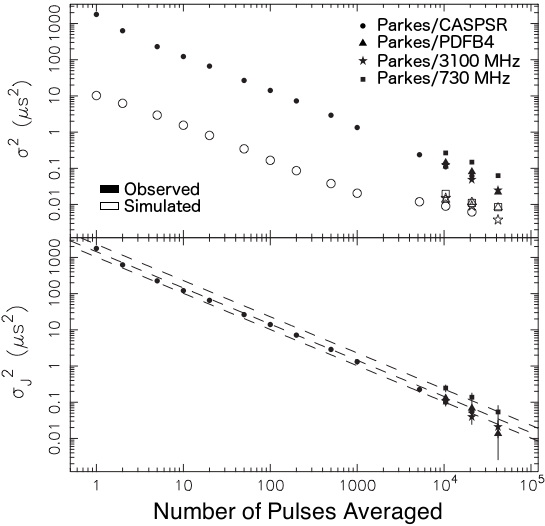}
\caption{Each rotation of a pulsar can present a vastly different profile (``jitter''), yet on timescales of $\gtrsim$1\,h the rotationally averaged profile is stable. For this reason, each timing residual data point is usually made up of an $\sim1\,$h pulsar observation. On the left, we show 30 single pulses from one of the best-timed pulsars. The top panel shows its stable profile after 1\,h of integration. The right panel, from \citet{shannonstefan+14}, shows the RMS residual level vs.\ number of pulses contributing to the integrated profile for real and simulated data (top), and the quadruature subtraction of these (bottom).}
\label{fig:avgpulse}
\end{figure*}

The basic data product of PTAs are ``timing residuals'', of which each data point is a pulse arrival time minus the pulse's predicted arrival time. %!!!Shown later in figs. x,y 
Each timing data point is built from a long observation of each pulsar---typically averaged over tens of thousands of rotations---long enough so that its profile is relatively stable and of a sufficiently high signal-to-noise ratio (Fig.\,\ref{fig:avgpulse}; see also \S\ref{sec:noise}). Typical durations are approximately 0.5--1\,h with current telescopes \citep{ppta-basics,nanograv-basics,epta}. The GW frequency sensitivity is set by the observing cadence, $C$, i.\,e.\ a pulsar typically gets one datum every $\sim$3\,weeks, and the total span of data, $T$, which can right now in principle be up to $\sim$30\,years, as precision timing was first done in the 1980s \citep{msp-timing}. Hence, $T^{-1}\lesssim f\lesssim C^{-1}$ gives the classically quoted \,nHz--$\,\mu$Hz sensitivity.

\begin{figure*}
\centering
\includegraphics[width=0.63\textwidth]{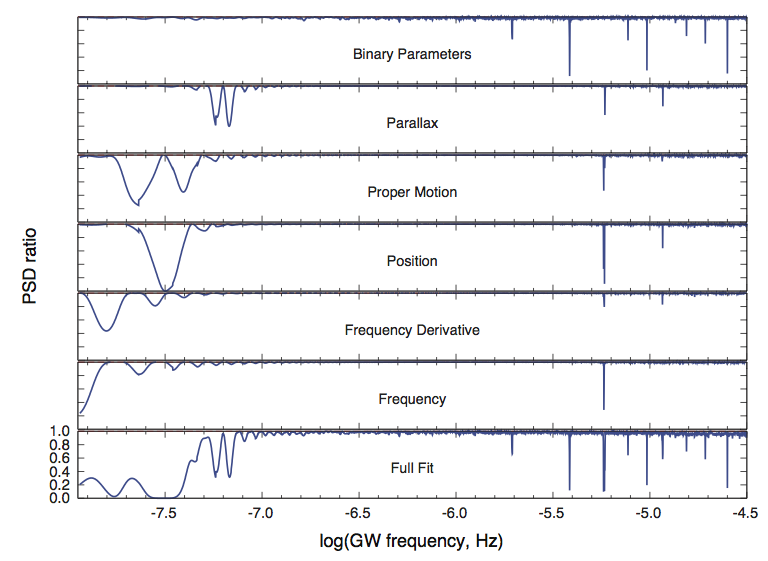}
\vspace{-3mm}
\caption{From \citet{cutler14}: GW power absorbed by parameter fitting for PSR\,J0613--0200. The ordinate is the ratio of power spectral density for residuals before and after parameter fitting. Only the indicated parameters were fit.
%Binary parameters will vary from pulsar to pulsar. 
Frequencies $\fgearth\sim1\,\yHz$ are fully absorbed by pulsar position fits, thus are not detectable by PTAs; this corresponds to the $1\,\yHz$ sensitivity spike of Fig.\,\ref{fig:overview}. PTAs are also decreasingly sensitive as $f\rightarrow\tobs^{-1}$, which in this simulation was just below the window shown here. 
%!!! MAKE AGAIN AT LOWER FREQUENCY WITH LONGER DATA SPAN? !!! 
(\citealt{blandford+84}; see also \citet{cordes13} for an excellent discussion of GW sensitivity limits due to intrinsic pulsar noise and parameter fitting. Copyright (2014) by The American Physical Society
}
\label{fig:absorption}
\end{figure*}

The predicted arrival time of a pulse is built from a model of the pulsar's spin evolution. Because intrinsic pulsar spin and binary properties are not directly measurable, these are solved for by the timing process itself: with any new data, the pulsar parameters are re-fit and a more accurate timing model is obtained before any GW search can be performed. Some GW power may be absorbed by this process (Fig.\,\ref{fig:absorption}). Fortunately, most intrinsic parameters vary from pulsar to pulsar, hence PTAs only suffer significant sensitivity losses at frequencies of 1\,year and 0.5\,year due to positional and parallax fitting, and at the lowest frequencies to period ($\propto t$) and period derivative fitting ($\propto t^2$).

A general figure of merit for pulsar timing precision is the root mean squared of the timing residuals, $\sn$, which for the best pulsars is $\lesssim$100\,ns, however is more typically $\lesssim$1\,$\mu$s for pulsars included in current PTAs.

\subsection{Gravitational waves' effect on pulsar signals}\label{sec:gws}
Gravitational radiation is a propagating change in the curvature of spacetime, parameterized by the dimensionless strain, $h$. Because we expect no Galactic source of gravitational waves to be large enough to be detected by PTAs,\footnote{Except, perhaps, for intermediate-mass black hole binaries in our Galaxy's globular clusters, \citep[\eg][]{globs}.} throughout this review we consider only GWs in the far-field regime: that is, well outside of our galaxy. From any GW source there is a path difference, and hence travel time delay, between the source-Earth path and the source-pulsar-Earth path. For a far-field source, a GW's appearance in timing residuals at time $t$ has been shown to be the difference $\Delta h_{+,\times} = h^p_{+,\times} - h^E_{+,\times}$ between the GW's distortion of Earth's space-time (``Earth term,'' $h^E_{+,\times}(t)$) and the GW's distortion of pulsar space-time (``pulsar term,'' $h^p_{+,\times}(t-d/c)$ for a pulsar of distance $d$). The plus and cross represent the two polarization states possible in general relativity.

\begin{figure*}
\centering
    \subfloat[Gravitational Wave Background\label{fig:classesgwb}]{%
        \includegraphics[trim=3.6cm 2mm 18mm 17mm,clip,width=0.49\textwidth]{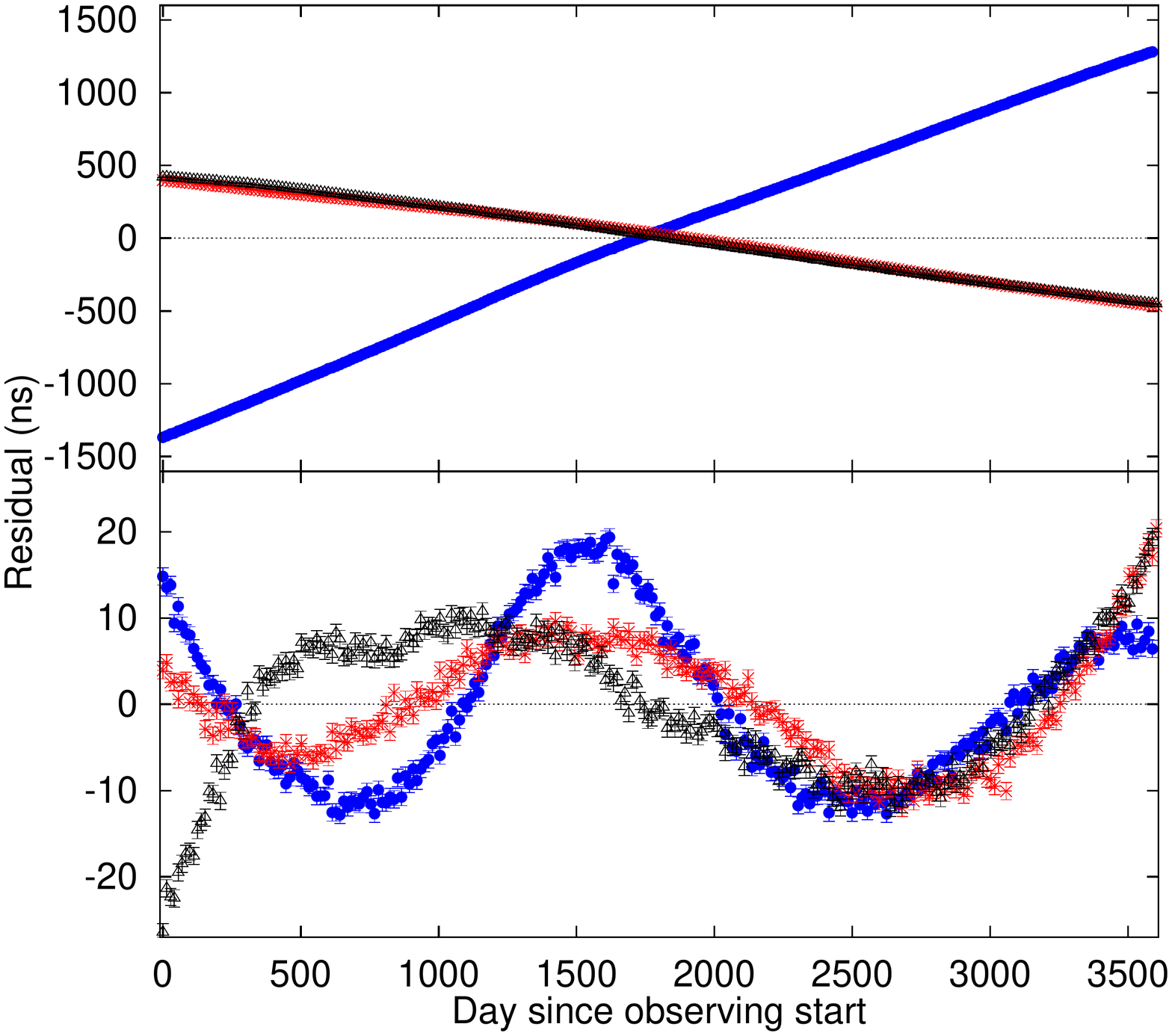}
    }
    \hfill
    \subfloat[Continuous Wave\label{fig:classescw}]{%
      \includegraphics[trim=3.6cm 2mm 18mm 17mm,clip,width=0.49\textwidth]{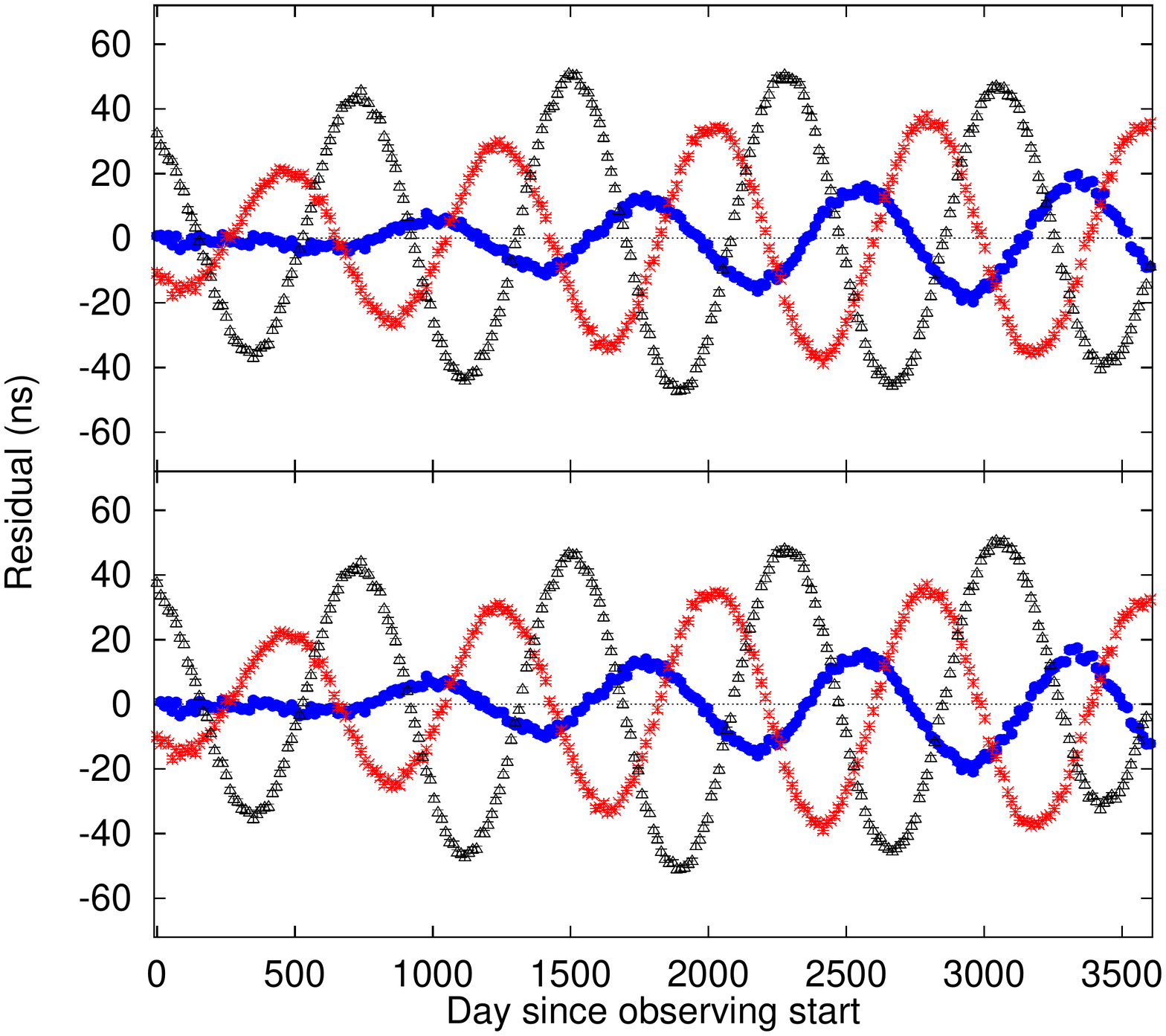}
    }\\
    \subfloat[Memory\label{fig:classesmemory}]{%
        \includegraphics[trim=3.6cm 2mm 18mm 17mm,clip,width=0.49\textwidth]{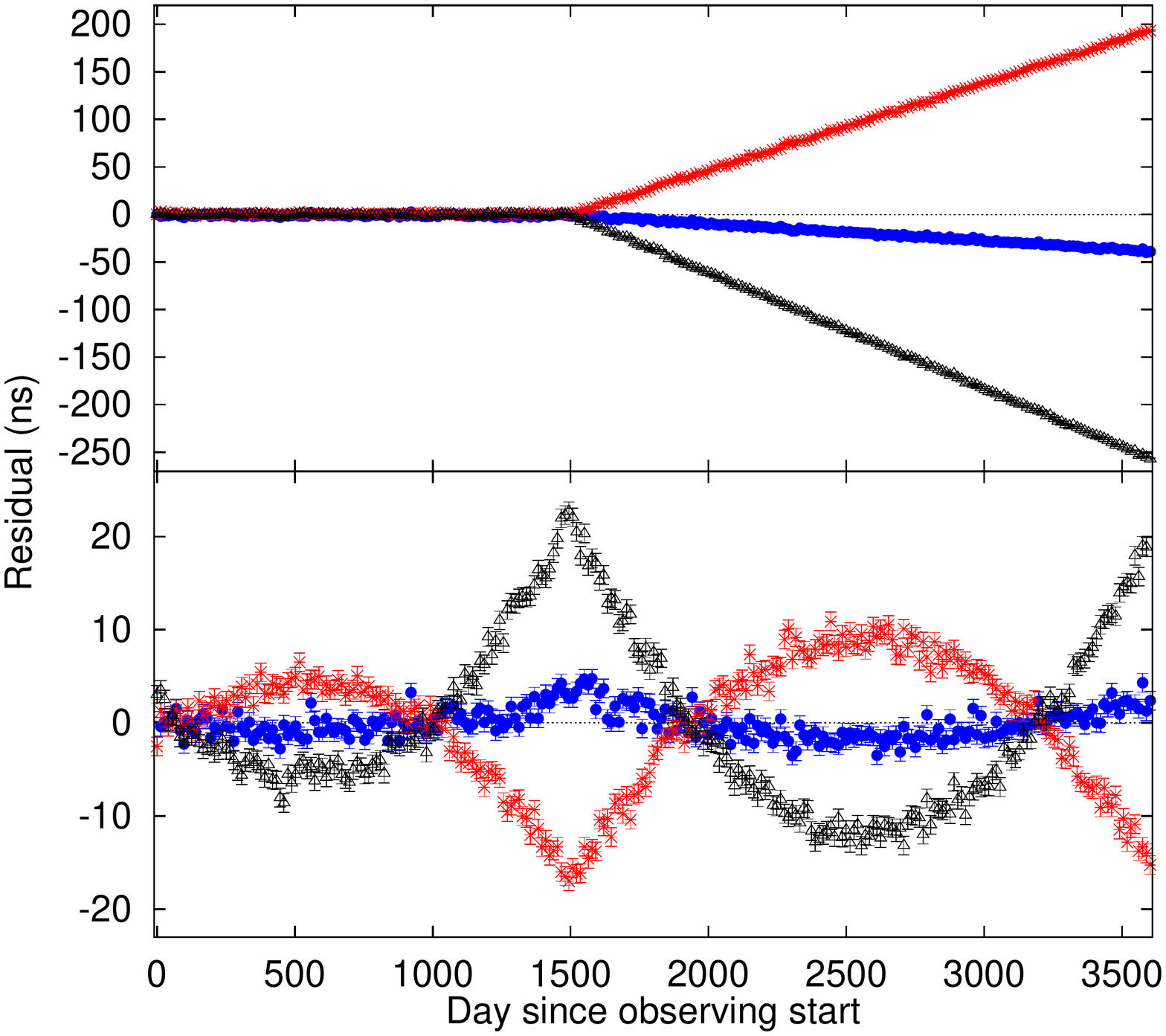}
    }
    \hfill
    \subfloat[Burst\label{fig:classesburst}]{%
      \includegraphics[trim=3.6cm 2mm 18mm 17mm,clip,width=0.49\textwidth]{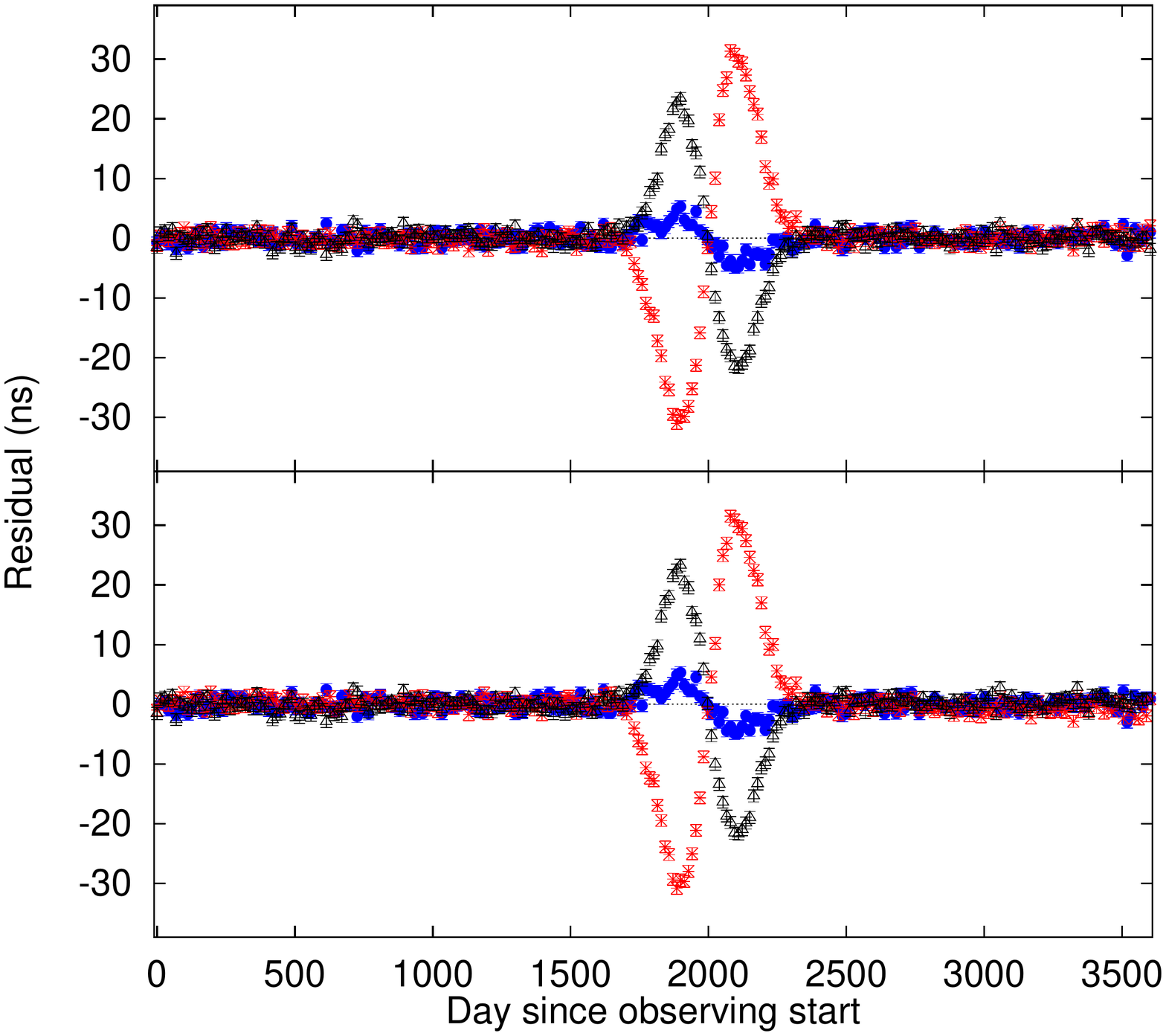}
    }
\caption{Examples of the four classes of GW signals predicted in the pulsar timing band. Each graphic shows the induced timing residuals before parameter fitting (top panel) and after fitting for pulsar spin period and period derivative (bottom panel), for simulated data from pulsars PSR\,J0437--4715 (red asterisks), J1012+5307 (blue dots), and J1713+0747 (black triangles) to demonstrate the expected quadrupolar signature. All discrete GW sources (b--d) were injected in the same sky location. In all simulated pulsar residuals, $\sigma_{\rm n}=1\,$ns of white noise and no red noise was injected. Panels are: \emph{(a)} A GWB with $\hc=10^{-15}$ and $\alpha=-2/3$; \emph{(b)} A continuous wave from an equal-mass $10^9\,\msun$ BSMBH at redshift $z=0.01$. The distortion from a perfect sinusoid is caused by the lower-frequency pulsar term. \emph{(c)} A memory event of $h=5\times10^{-15}$, whose wavefront passes the Earth on day 1500. \emph{(d)} A burst source with an arbitrary waveform. 
%and simulates the induced signal in multiple pulsars to demonstrate the expected correlations. In all cases we have simulated a 20\,year data set and negligible intrinsic pulsar noise.(!!!)
}
\label{fig:classes}
\end{figure*}

As in \citet{detweiler}, we can phrase the GW-induced residual signal as a shift in the pulsar spin frequency, $\nu_0$.\footnote{Spin frequency is constant here, which is sufficiently true after correcting for the fitted pulsar spin derivative and parameters.} For a GW travelling in cartesian coordinates along the positive $z$ direction, the induced frequency shift depends on a pulsar's direction cosines $\alpha, \beta$, and $\gamma$ to the $x, y$, and $z$ axes. The coefficients in front of the cross and plus terms in the equation below define the quadrupolar ``antenna pattern'' expected to be seen from GWs when correlating the residuals of pulsars with different direction cosines.
The frequency shift is given by
\begin{equation}\label{eq:redshift}
   \frac{\nu_0 - \nu(t)}{\nu_0} = \frac{\alpha^2 - \beta^2}{2(1+\gamma)}\Delta h_{+}(t) + \frac{\alpha\,\beta}{1+\gamma}\Delta h_{\times}(t)~.
\end{equation}
Finally, the residuals induced by the GW are given by the integral over time of Eq.\,\ref{eq:redshift}:
\begin{equation}
   R(t) = \int^t_0 \frac{\nu_0 - \nu(t')}{\nu_0}\delta t'~.
\end{equation}
Fig\,\ref{fig:classes} shows examples of $R(t)$ for several source classes.

%Thus, a distant source will produce distortions at the Earth and at the pulsar that are observed from earth to be different (either delayed in time for a transient GW, or simply different in frequency for a frequency-evolving GW source like an inspiralling binary supermassive black hole).

%Important things to include:
%\begin{itemize}
%\item What is a “residual”?
%\item Mention the use of MSPs
%\item Pulsar term vs.\ earth term
%\item PTAs can detect $T^{-1}\lesssim f\lesssim C^{-1}$ although the actual sensitivitiy is a little less.
%\end{itemize}

%\noindent (
% --$>$ A review to summarise the current status, sensitivities, detection goals, and anticipated future of PTA detection of GWs.
% --$>$ Basics e.g. determining strain from residuals.
% --$>$ Will include relevant maths as required by the other sections.)

\subsection{Leading PTA experiments}\label{sec:ptas}
The sensitivity scaling of PTAs depends on the type of target signal, however the factors contributing are minimizing the pulsars' $\sn$ values, and the presence of ``intrinsic red noise'' in a pulsar's timing residuals (\S\ref{sec:noise}), while maximizing the total number of pulsars, the observing cadence, and the total data span. The most sensitive PTA experiments optimize these aspects.
%Pulsar timing dates back to the first discovery of pulsars. 
%Table \ref{table:ptas} gives a summary of notable timing programs.
In the past decade, there has been a push towards large, coordinated pulsar timing programs. We will describe the leading programs here in brief; please see the noted publications for details.

%While scattered efforts in pulsar timing date back to the first time the technique was proposed citep{who-was-that?}, there has been a recent push towards large coordinated programs. We will describe these programs here in brief; please see the noted publications for details.

{\bf The North American Nano\-hertz Observatory for Gravitational Waves} (NANOGrav)\footnote{http://nanograv.org/} has been running for $\sim$1 decade \citep{9yrNANOdata}, and is currently using sensitive wide-bandwidth receivers at Green Bank and Arecibo telescopes to time 49 pulsars (and counting; P.\ Demorest, private comm.), each observed at least at two radio frequencies between 327\,MHz and 2.3\,GHz. The reason for bi-frequency timing is to mitigate effects of the inter-stellar medium on the timing residuals; this is discussed further in \S\ref{sec:noise}. Arecibo telescope has the benefit of the largest collecting area---and largest sensitivity---of all the world's telescopes, although its size inhibits its sky coverage, so that it only may observe 27 of the current NANOGrav pulsars.

{\bf The European Pulsar Timing Array} (EPTA)\footnote{http://www.epta.eu.org/} is a collaboration which uses five telescopes across Europe---Effelsberg, Lovell, Nan\c{c}ay, Sardinia, and Westerbork---to separately time a primary set of 41 pulsars at various frequency pairs between 300\,MHz and 3.5\,GHz (\citealt{epta}; Desvignes in prep.). The collaboration is also working to combine all telescope signals coherently in a ``Large European Array for Pulsars,'' which will effectively be a fully-steerable telescope with a collecting area equal to that of Arecibo \citep{epta}. Because five telescopes contribute independently to EPTA data, it boasts the highest cadence of the three PTAs, effectively having one data point per pulsar each week.

{\bf The Parkes Pulsar Timing Array} (PPTA)\footnote{http://www.atnf.csiro.au/research/pulsar/ppta/} times 25 pulsars using Parkes Observatory in Australia at frequencies of 600\,MHz, 1.5\,GHz, and 3\,GHz. This experiment is unique due to the telescope's position in the southern hemisphere; it can see a distinct set of pulsars not visible to the EPTA nor NANOGrav, which includes one of the best-timed pulsars in the sky, PSR\,J0437--4715 \citep{georgeIPTA}. The PPTA has continually boasted the best constraints on gravitational radiation.
%, in part because it includes three powerhouse pulsars with the lowest measured RMS residuals: PSR\,J1909--3744, J0437--4715, and J1713+0747.

Finally, the aptly-named {\bf International Pulsar Timing Array} (IPTA)\footnote{http://www.ipta4gw.org/} represents the coordinated effort to combine data from the above three projects to reach the best possible sensitivity to GWs \citep{georgeIPTA,dickIPTA,mauraIPTA}. For its initial 2015 data release, %\citet{jorisinprep}, 
the array contains a total of 49 pulsars (Verbiest et al.\ in prep). Note this number is not the sum of the above collaborations' pulsars because the PTAs do not time mutually exclusive pulsars, and the 2015 IPTA data release does not contain pulsars that have been added to the constituent arrays in only the last few years. In addition to coordinating data and combined-PTA science, the IPTA hosts annual PTA science meetings around the world, and organizes related public efforts like PTA Mock Data Challenges \citep[\eg][]{mock-data-challenge1,mock-data-challenge2,mock-data-challenge3}.

%So that we can later quote their sensitivities. Make it brief
%\noindent --$>$ NANOGrav, PPTA, EPTA, SKA-PTA, Chinese PTA.

\section{Upper Limits and Detection of GWs}\label{sec:gwsrc}
There are four distinct ``classes'' of signals which might appear in the nHz-$\mu$Hz GW band, with several astrophysical sources that can produce them. A simulated example of how each might appear in timing residuals is shown in Fig.\,\ref{fig:classes}.

Different techniques have been developed to place upper limits on each class of signal. However, for all classes, the only way to verify a detection is to demonstrate the expected quadrupolar signature, for instance via the so-called ``Hellings and Downs curve,'' which demonstrates the expected cross-correlation function for two pulsars as a function of their angular separation on the sky \citep[][]{hellings-downs,hellings-downs-finn}.\footnote{This curve has been shown to hold both for a stochastic background, and for a discrete source of GWs. However, the expected overlap reduction function may differ for non-general-relativistic theories of gravity (\eg\ \citealt{non-gr-hd}). The pulsar term (\S\ref{sec:gws}) will also cause scatter around the predicted curve.} For this reason, while upper limits can be placed using data from one pulsar, \emph{at least three pulsars} must be used in an array to demonstrate that a GW has been \emph{detected}.

%Below we describe each class of signal, note sources that might cause it, review methods for detection/upper limit, and note the limits placed to date on these classes.

\begin{figure}[h]
\centering
\vspace{-4mm}
\includegraphics[width=1.0\columnwidth]{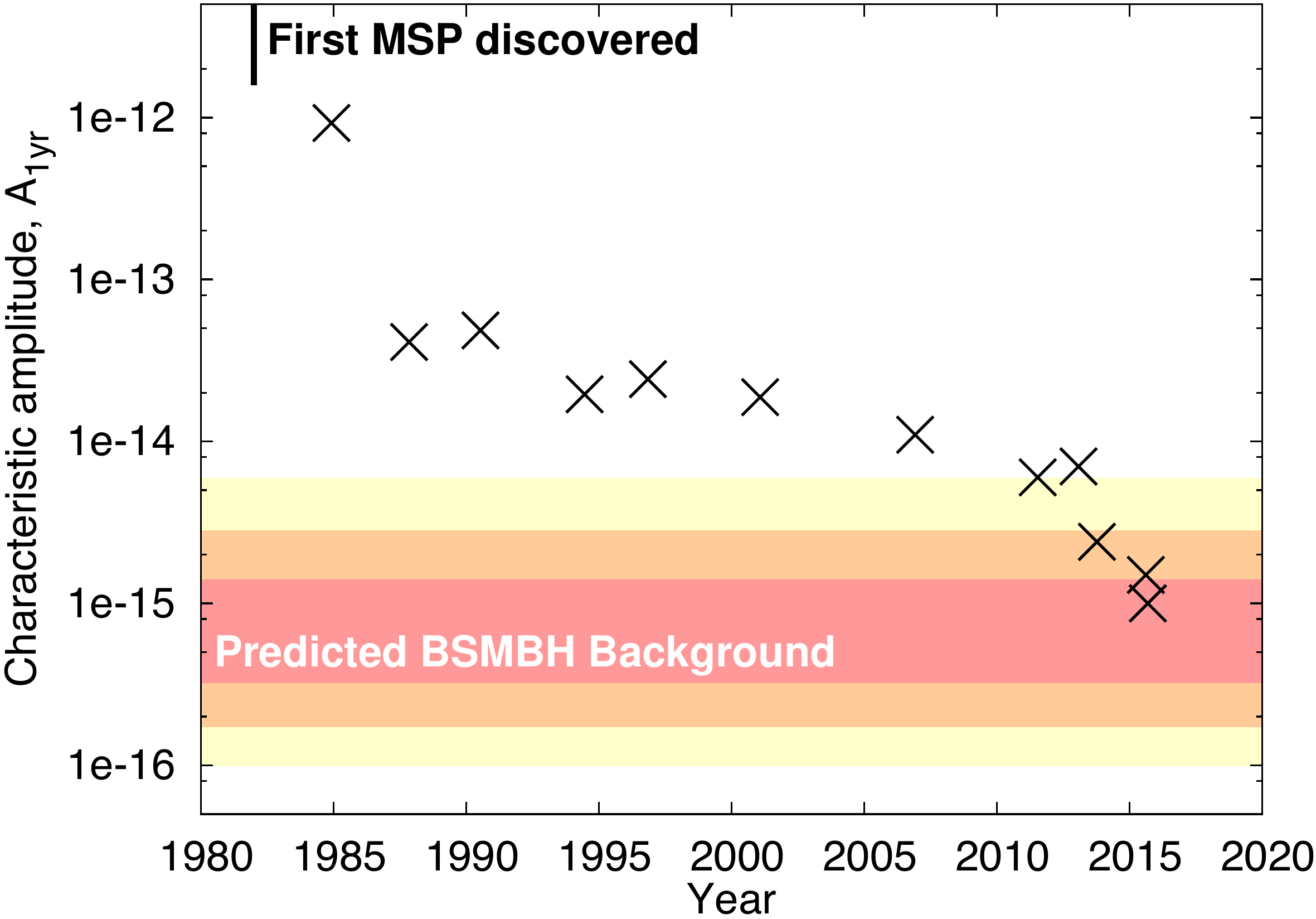}
\vspace{-6mm}
\caption{Upper limits on the power-law GWB for a spectral index $\alpha=-2/3$. Limits improved steadily after dedicated timing of millisecond pulsars commenced. The sudden drop after 2013 arises, most importantly, from optimized PTA experiments finally coming to fruition. The red, orange, and yellow ranges give the 68, 95 and 99.7\% confidence intervals of the \citet{sesana13} model. 
{\scriptsize References: \citet{first-msp,blandford+84,rawley+87,stinebring+90,kaspi+94,mchugh+96,lommen-thesis,jenet+06,vanhaasteren+11,demorest+13,pptascience1,9yrNANOlimit,pptascience2,lentati+15}}}
\label{fig:avstime}
\vspace{-4mm}
\end{figure}

\subsection{Stochastic Gravitational-Wave Background (GWB)}\label{sec:gwb}
This describes a signal built from the ensemble contributions of discrete GW sources. It manifests as randomly varying strain fluctuations with a well-defined spectral distribution. The GWB's spectrum, as a fractional contribution of the GWB to the energy density of the Universe in a logarithmic frequency interval, 
% in a logarithmic frequency interval
can be expressed (assuming a Freedman-Robertson-Walker universe) as
\begin{equation}
%\Omega_{\rm gw}(f) = \frac{2\pi^2}{2H^2_0}f^3S_h(f) ; S_h(f) = \frac{\hc(f)^2}{f}\\
\Omega_{\rm gw}(f) = \frac{2\pi^2}{3H^2_0}f^2h_c(f)^2~,
\end{equation}
\citep[\eg][]{maggiore00}, where $H_0$ is the Hubble constant, and $h_c(f)$ represents the ``characteristic strain spectrum,'' which is built from the quadrature sum over the individual strains from a GW source population.

For a time, most predictions of the GWB parameterized the strain spectrum by a power law, such that the strength of the background can be characterized by an amplitude $\ayr$ for an index $\alpha$,
\begin{equation}\label{eq:powerlaw}
h_c(f) = \ayr\Bigg(\frac{f}{1\,{\rm yr}}\Bigg)^\alpha.
\end{equation}
Pulsar timing limits are often quoted in one of three ways: either the $h_c$ or $\Omega_{\rm gw}$ limit at $f$ where $f\gtrsim T^{-1}$, or as a limit on $\ayr$ for a specific $\alpha$ value. The most sensitive limits are placed at $f > T^{-1}$ because of signal absorption by pulsar fitting at $f\lesssim T^{-1}$ (Fig.\,\ref{fig:absorption}).
%\citep[\eg][]{jenet+06,demorest+13,pptascience1}. 
Reporting an ($\ayr$,$\alpha$) pair allows easy comparison between upper limits from different PTA experiments.

Current consensus is that the dominant GWB signal in the nanohertz waveband will be from binary supermassive black holes (BSMBHs; where $\alpha=-2/3$ \citep{phinney01}. See also \S\ref{sec:smbhunknowns}), however predictions have also been made for cosmic strings ($\alpha=-5/3, -7/6,$ or $-1$ depending on the frequency range and kink vs.\ cusp signal; \citealt{damourvilenkin01,olmez+10}), and inflationary relic GWs ($\alpha=-2$ to $-0.5$ depending on the equation of state during the inflationary epoch, \eg\ \citealt{grishchuk05}). Other suggested sources of a GWB, such as primordial black holes \citep{bugaev+11} and QCD phase transitions \citep{caprini+10}, do not fit a power-law dependence for $\hc$. All predicted PTA backgrounds have greater power at low frequency. The result of this is a red-noise wander in the timing residuals that is correlated across different pulsars, as seen in Fig.~\ref{fig:classesgwb}.

%\subsubsection{GWB Limits}
%
%Most algorithms for placing limits on the GWB generally involve 
%Make residuals
%Model pulsar noise
%Do something to the data e.g. whiten and form power spectra 
%Set a detection statistic
%(Frequentist) Inject a GWB into fake data with pulsar-like noise, and adjust the GWB to a strength at which in 95\% of the  simulations, the detection statistic exceeds the observed detection statistic. 
%(Bayesian) Doing bayes magic
%modelling the noise in each pulsar (including a red and white component; see \S\ref{sec:noise}), and then determining how significant a background could be before it would dominate the noise.

We will not review limit algorithms, however do encourage readers to read about frequentist approaches in \eg\ \citet{pptascience1,ellis13freqBG}. Bayesian inference is also increasingly used, \eg\ \citet{vanhaasteren+11,ellis13bayesBG,lentati+15}.

As of yet there has been no detection, however increasingly tight upper limits have been placed on the GWB. Upper limits to date for $\ayr$ and $\alpha=-2/3$ are shown in Fig.\,\ref{fig:avstime}, indicating a steady improvement with a drop in nearly an order of magnitude over the last decade. See also \S\ref{sec:smbhunknowns}, and Figs.\,\ref{fig:turnovers} and \ref{fig:astrophysics} for other views and further implications of PTA limits for the BSMBH population.

The best limits to date on other sources include \textbf{cosmic strings}, \mbox{$\ayr<6\times10^{-16}$} at \mbox{$\alpha=-7/6$} \citep{9yrNANOlimit}. This limit's interpretation depends heavily on a number of parameters, including string loop radius and the probability of two strings connecting; see \eg\ Fig.\,12 in \citet{9yrNANOlimit}. However, for the range of parameters assumed by Arzoumanian et al., they set a conservative upper limit on the string tension of $G\mu<3.3\times10^{-8}$, which is four times more constraining than previously published limits. See also \citet{lentati+15} for an excellent discussion of the impact of PTA limits on cosmic strings. Current limits also impact \textbf{inflationary GWs}, \mbox{$\ayr<8.1\times10^{-16}$} at \mbox{$\alpha=-1$}. Following the framework of \citet{zhao11}, this provides a limit on the Hubble parameter during the inflationary era of \mbox{$H_*/m_{\rm P}<1.6\times10^{-2}m_{\rm P}$}, where $m_{\rm P}$ is the Planck mass \citep{9yrNANOlimit}.
% Find somewhere to write in more detail about why cosmic strings and inflation are important?

\begin{figure}
\centering
\vspace{-1mm}
\includegraphics[width=0.85\columnwidth]{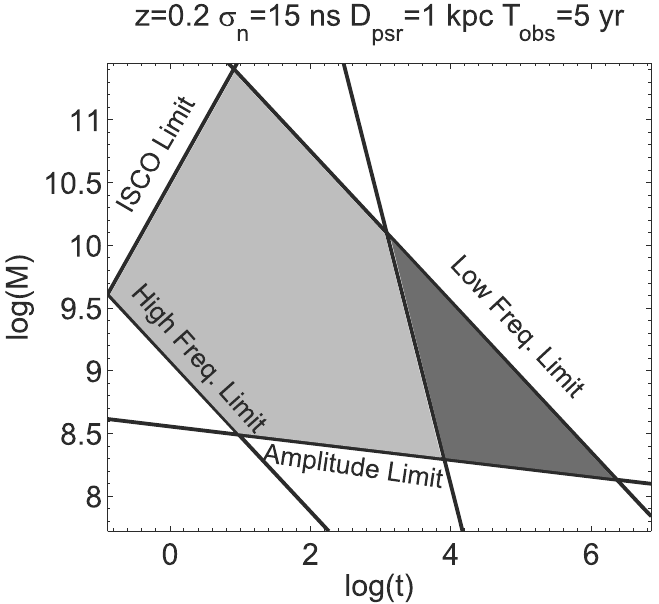}
\vspace{-1mm}
\caption{From \citet{lee+11}: The parameter space of BSMBHs detectable as CW sources by a representative PTA. The abscissa is log($t_{\rm gw}$) (Eq.\,\ref{eq:tgw}) in years; the ordinate is the chirp mass in units of $\msun$. The ``Amplitude Limit'' depends on PTA sensitivity, and the ``Low and High $f$ Limits'' bound the PTA GW band.
``ISCO'' is the inner-most stable circular orbit. Evolving and non-evolving binaries are light and dark grey, respectively. The dominantly detectable population is in the dark gray region, where we encounter
smaller masses but much longer system lifetimes.
}
\vspace{-0mm}
\label{fig:keija}
\end{figure}

\begin{figure}
\centering
\hspace{-5mm}\includegraphics[width=0.9\columnwidth]{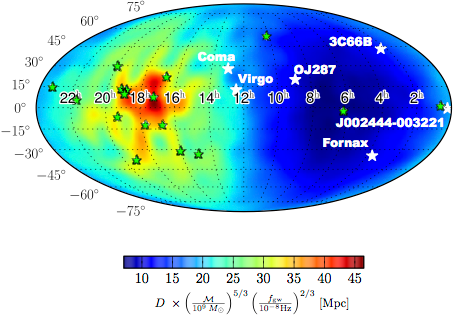}\\
\includegraphics[width=0.85\columnwidth]{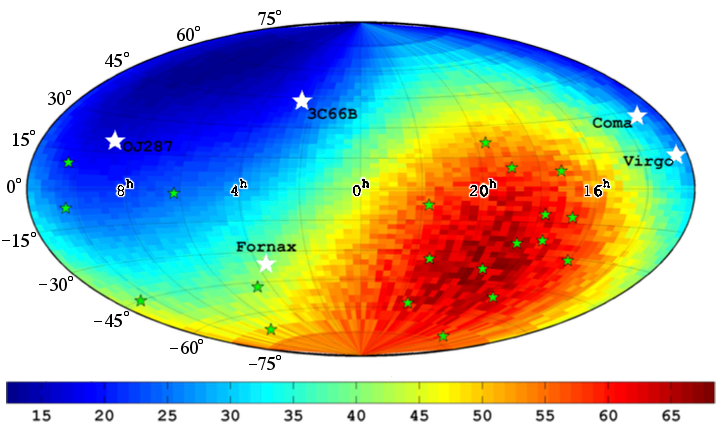}
\caption{From NANOGrav \citep[top,][]{nano-cw}, and PPTA \citep[bottom,][]{ppta-cw}: At the time of those publictions, the respective PTAs were sensitive to SMBHs out to the color-coded distance $D$, normalized here for an equal-mass $10^{9}\,\msun$ binary with $\fgearth=10^{-8}$\,Hz. Note the 12\,h R.A.\ shift between the panels. The difference in sky sensitivity distribution is caused by the different positions of the pulsars in the arrays (small green stars).}
\vspace{-3mm}
\label{fig:cwskyplot}
\end{figure}

\subsection{Continuous Waves (CW)}\label{sec:cw}
CWs are, as their name implies, continuous in time. The generation of bright CWs in the nHz--$\mu$Hz GW band requires massive objects in a state of constant acceleration at lengths of decades; currently the only sources known to fit this requirement are BSMBHs whose orbit is gradually shrinking due to GW emission. If a binary in its rest frame is orbiting at a frequency $\fbrest$, a circular binary will emit GWs at a rest-frame frequency $\fgrest=2\fbrest$, and an Earth-observed frequency $\fgearth=2\fbrest/(1+z)$. Any eccentricity in the system will spread the observed GW power to higher harmonics of $\fgearth$ \citep[\eg][]{enoki+07}. Thus, binaries of orbital periods in the range of weeks to decades will emit CWs in the PTA band. These systems spend more time at longer orbital periods, as per the timescale to coalescence given by a purely GW-driven orbital inspiral \citep{milomerrittLRR}:
\begin{equation}\label{eq:tgw}
%t_{\rm gw} = \frac{5}{256F(e)}\frac{c^5}{G^3}\frac{a^4}{\mu M^2},
t_{\rm gw} = \frac{5}{256F(e)}\frac{c^5}{G^{5/3}}\frac{(m_1+m_2)^{1/3}}{m_1m_2}(\pi\fgrest)^{-8/3},
%m_1m_2(m_1+m_2)},
\end{equation}
where $a$ is the semi-major axis of the system and $m1$, $m2$ are the masses of the two black holes. Because of the time delay between the Earth and pulsar term can be several tens of thousands of years (see \S\ref{sec:basics}), there may be some evolution in the BSMBH between these two terms. This would cause two sinusoids of different phase and frequency to appear in each pulsar's data stream: a low-frequency pulsar term and a higher-frequency Earth term, the latter of which would be the same in all pulsars in the PTA. A source with a significant difference between Earth and pulsar term is said to be ``evolving''. Figure \ref{fig:keija} indicates the parameters defining what BSMBH systems would be expected to emit CWs detectable by a typical PTA experiment. The signal from a weakly evolving source is shown in Fig.\,\ref{fig:classescw}.

The GWB is generally expected to be the dominant signal in the PTA band, however numerous studies have indicated that CW signals from BSMBHs will be resolvable above the stochastic background \citep[\eg][]{sesanasingle,boylepen12,rosado-gwb}.
A number of both targeted and blind searches have been performed for CWs from circular BSMBHs. These searches are able to translate GW upper limits into statements on what BSMBHs must not exist out to some distance at a given mass, mass ratio, and orbital frequency. Targeted limits have been placed on binaries in SgrA* and nearby galaxies \citep{lommenbacker01}, and perhaps the most noteworthy targeted limit was that on GWs from 3C66B, which had been suggested via an observation of elliptical motion of its radio core to be a BSMBH system \citep{sudou3C66B}. The implied system parameters for 3C66B were quickly ruled out because the corresponding GW signal would have been detected by the PPTA with great significance \citep{jenet3C66B}.
To date the most sensitive upper limits from blind BSMBH searches are shown in Fig.\,\ref{fig:cwskyplot}, along with several nearby clusters and candidate BSMBHs. 
%!!! See also Fig.\,\ref{fig:mm}

\subsection{Memory Events}\label{sec:memory}
GW memory describes an event which causes a static (non-oscillating), propagating change in strain. This is expected to occur upon the coalescence of a BSMBH, or from a BSMBH on a hyperbolic orbit \citep{favata09a,favata09b}. Physically, GW memory passing Earth can be thought of as a permanent displacement of the local space-time following a rapid ($\ll1\,$day) ramp in the local metric as the wave passes. The ramp itself is not detectable by pulsar timing, but the sudden change in Earth's space-time induces a small change observed in the pulsar period, which results in a gradual drift of timing residuals away from null because the previous timing model is rendered inaccurate (Fig.\,\ref{fig:classesmemory}).

The memory signal would also appear in the pulsar term, however it would not be coincident in time with other pulsars: the pulsar term is detected at a time $(d/c)(1+cos(\theta))$ later than the correlated Earth term, where $\theta$ is the separation angle between the pulsar and the GW source as seen from the Earth.

Predictive simulations have found that PTAs are highly unlikely to detect GW memory, largely due to the extreme rarity of bright events \citep{seto09,rutgermem,cordesjenet12}. Still, techniques to search for memory in PTAs have been developed \citep{madisonmem} and limits have been placed: most recently, results from PPTA and NANOGrav bounded the event rate of events with strain $h>10^{−13}$ to be $<$0.75\,yr$^{-1}$ and $<$1.5\,yr$^{-1}$, respectively \citep{ppta-mem,nano-mem}. However, PTA sensitivity is still several orders of magnitude above the predicted memory strain signals \citep{ppta-mem}.

\subsection{Transient Bursts}\label{sec:burst}
For PTAs, a transient or burst source is defined as an event whose duration $\tau\ll T$; and of course it would only be detectable if $\tau\gtrsim 1\,$day, such that the residuals of multiple pulsars could be observed to be affected by the burst signal. 

Bursts may arise from cusps and kinks in cosmic strings \citep{damourvilenkin01}, or from encounters of compact objects with a SMBH, including SMBHs in an unbound orbit \citep{finnlommen10}. 
While several methodologies for burst detection via PTAs have been developed \citet{finnlommen10,deng14,zhu-generic}, as of yet none have been published which apply the techniques to real (non-simulated) PTA data.

%Add the sensitivity scalings of each of these? e.g. GWB from jenet+06, memory from van haasteren & levin, etc...

%In Figure \ref{fig:strainspec}, we show the PTA strain spectrum in the context of other detectors, and also demonstrate the relative anticipated strength of various GW sources in the PTA band. It is clear here that by most predictions, GWs from binary SMBHs dominate other signal sources, although the range of predictions for the cosmic string background vary enormously. Here, we give an overview of the work to date on predicting the strength of various GW signals. In the remainder of the manuscript, we focus on SMBH binaries, which have had the greatest

\section{Current discussions and challenges in pulsar timing}
%The efforts involved in GW detection with PTAs are broad-ranging
Here we touch on major efforts that aim to maximize PTA sensitivity and understand the future of GW detection with PTAs.
One further major effort is in the modelling of BSMBH populations; however, this is covered in detail in \S\ref{sec:smbhunknowns}.

\begin{figure*}
\centering
\includegraphics[trim=2.3cm 4cm 1.5cm 3cm,clip,width=0.7\textwidth]{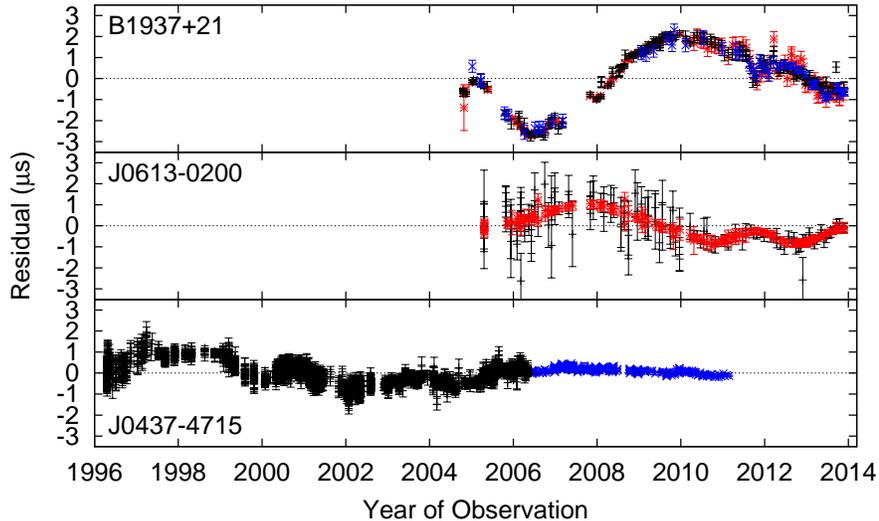}
\caption{Timing residuals from three pulsars, exhibiting different kinds of noise: intrinsic red timing noise (PSR\,B1937+21), a S/N limited pulsar with a mild level of red timing noise (PSR\,J0613--0200), and a pulsar exhibiting uncorrected dispersion measure variations (PSR\,J0437--4715 prior to 2007 was only observed at a single frequency, thus is not corrected for this effect prior to that date). Colors indicate the radio frequency of the observation: $\sim$1\,GHz (black), $\sim$800\,MHz (red), and $\sim$3\,GHz (blue). Data are from NANOGrav \citep[PSRs\,B1937+21, J0613--0200: ][]{9yrNANOdata} and the PPTA \citep[PSR\,J0437--4715:][]{ppta-basics}. See \S\ref{sec:noise} for discussion.}
\label{fig:residuals}
\end{figure*}

\subsection{Conquering timing noise}\label{sec:noise}
%\citet{jenet+06} found that for a BSMBH background, the sensitivity scaling of a PTA's sensitivity to $\ayr$ is
%\begin{equation}\label{eq:scaling}
%\ayr \propto \sn T^{-5/3} \sqrt{N_{\rm pts} N_{\rm psr}}~,
%\end{equation}
%where $N_{\rm pts}$ is the total number of data points per pulsar, and $N_{\rm psr}$ is the number of pulsars in the array.
The RMS level $\sn$ of timing redsiduals is critically important to PTA sensitivity, and for many pulsars encompasses both a red (increasing at low power) and white (gaussian) noise component. Examples of timing residuals dominated by different types of noise can be seen in Fig.\,\ref{fig:residuals}.

Some noise sources are surmountable. For most pulsars observed with current telescopes, the dominant white noise contribution comes from the low signal-to-noise (S/N) of the observed profile used to make a timing residual.
Such pulsars are ``S/N limited''. This is perhaps the most (conceptually) easy noise to overcome, simply requiring a more sensitive radio telescope, which would involve recording at higher bandwidth \mbox{(${\rm S/N}\propto\sqrt{B}$)}, better cooling one's receivers, or having greater collecting area \mbox{(${\rm S/N}\propto A$)}. Improvements from higher bandwidth are visible in PSR\,J0613--0200 in Fig.\,\ref{fig:residuals}; in 2010, the new ``GUPPI'' instrument was installed at Greenbank telescope, which raised the bandwidth that the instrument could observe by more than a factor of 10. As discussed later, however, even S/N improvements cannot overcome intrinsic pulse ``jitter.''

One of the main sources of red noise in timing residuals comes from turbulence in the interstellar medium (ISM). As the pulsar, ISM, and Earth itself move, we see pulsars through a continuously changing line-of-sight through the ISM, and the resulting electron density variance creates a wandering delay in the pulse's arrival at Earth (``dispersion,'' \eg\ \citealt{lorimerkramer}). Fortunately, this effect is radio-frequency-dependent and can be directly measured using at least two broadly-separated radio bands; this is the primary reason for multi-frequency observations noted in \S\ref{sec:ptas}. For PSR\,J0437--4715, the PPTA only commenced multi-frequency observations in 2006; the strong red signal diminishes thereafter due to the correction of dispersion variations (Fig.\,\ref{fig:residuals}). The remaining low-level signal is likely due to intrinsic timing noise in this pulsar.
%!!! It has also been proposed that a separate telescope with a wide field of view can efficiently monitor PTA pulsars explicitly to measure and remove dispersion variations in the timing residuals measured at high frequency \citep[\eg][]{lofar-msps}.
%\footnote{Pulse scattering effects that are severe at low frequency (timescales much greater than a millisecond pulsar's rotational period) prevent precision timing at low frequencies.}

%People rectify this by measuring at two bands or having another telescope (like LOFAR or LWA---look this up) track DM variations.

Not all pulsars appear to be limited by extrinsic noise sources: for instance greater receiver bandwidth after the year 2010, and hence greater S/N, did not improve PSR\,B1937+21. All three pulsars shown in Fig.\,\ref{fig:residuals} retain some red noise signal despite dispersion corrections. The remaining noise is considered intrinsic noise, which is likely due to short (``jitter,'' Fig.\,\ref{fig:avgpulse}) and long-term rotational instabilities in the pulsar itself \citep{cordes13}.

This is a critical notion for the long-term sensitivity of pulsar timing and in designing future experiments: wherein even using ideal telescopes with maximized sensitivity, a fundamental instability in each of our pulsars could be a basic impediment to GW detection.
The pulsar emission mechanism is a vast field of pursuit and it seems that intrinsic pulsar noise is non-deterministic \citep{rickett75}. Thus, a number of current studies have aimed to find practical ways to side-step jitter, for instance creating timing residuals using pulse profiles made up of only high S/N rotations \citep{stefan+11,shannonstefan+14}. However, so far the only tried and true way to mitigate jitter is to integrate over a large number of pulses ($\sn\propto N^{-1/2}$; \citealt{shannoncordes1713}). Attempts at GW detection/limits, particularly those in pursuit of the GWB, must also account for intrinsic red signals present in pulsars \citep[\eg\ by noise modelling, by selecting only white-noise-dominated pulsars, and by ``pre-whitening'' schemes;][]{pptascience1}.

Finally, it is possible that other issues, such as time-transfer from atomic clocks or our knowledge of solar system ephemerides, will form a fundamental sensitivity limit for PTAs \citep{jenet-fundamental-limit}. However, because of constant ongoing improvements in terrestrial clock precision and solar system ephemerides, at least in the coming decades these sources of noise are not expected to dominate timing error. 

%The study of jitter is a big thing, with people attempting to understand whether, for instance, selecting only high S/N rotations to contribute to the net profile would result in improvements in S/N \citep{shannon,cordes,maybeStefan}. Other efforts have taken the challenge to understand why, fundamentally, a pulsar has a noise floor, although results are not yet conclusive \citep{}.

%\begin{figure}
%\centering
%\includegraphics[]{summary_plot_final.pdf}\\
%\caption{From \citet{taylor+submitted}: The probability of detection vs.\ year for NANOGrav, for BSMBH backgrounds of strength $\ayr=5.6\times10^{-16}$ (dash-dot), $\ayr=1\times10^{-15}$ (dashed), and $\ayr=2\times10^{-15}$ (solid). The three panels show red intrinsic timing noise that induces a residual RMS of 0, 5, and 10\,ns at 5\,y. This simulation used a realistic scenario for $\sn$ and number of contributing pulsars, taking actual NANOGrav $\sn$ values and adding three new pulsars per year at $\sn=200$\,ns, which is the median $\sn$ of the 17 pulsars in \citep{demorest+13}.}
%\label{fig:timetodet}
%\end{figure}

\begin{figure}
\centering
\includegraphics[width=0.95\columnwidth]{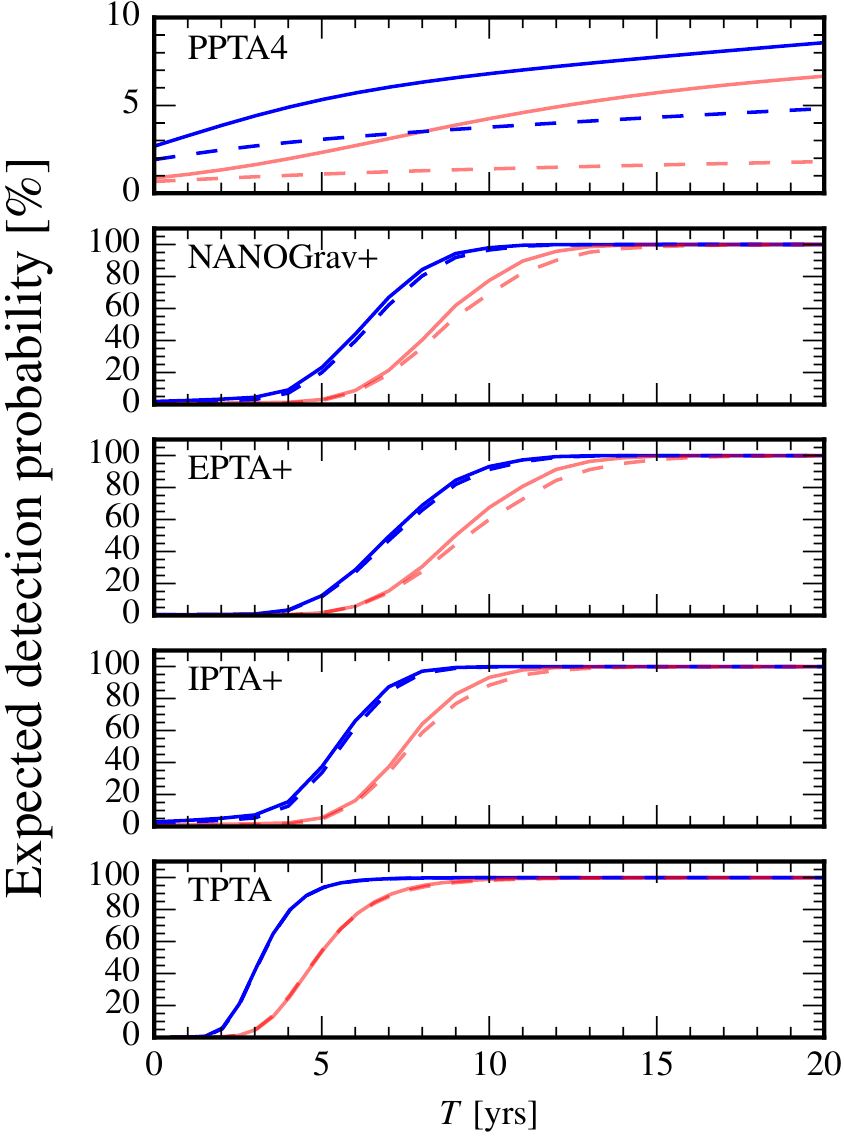}\\
\caption{From \citet{taylor+submitted}: The probability of detection of a GWB vs.\ further observing time for a PTA made up of the best 4 pulsars of the PPTA \citep{pptascience2}, an augmented NANOGrav (current array plus 4 new $\sn=250\,$ns pulsars per year), an augmented IPTA (6 new $\sn=250\,$ns pulsars per year), and a theoretical PTA, consisting of 50 pulsars each with $\sn = 100$\,ns and no intrinsic timing noise. The GWB amplitude distribution of \citet{sesana13} was used, on top of which various astrophysical effects were considered: blue and red lines show zero and 90\% of binaries stalled, respectively. Dashed lines correspond to the case where the true background spectrum has a turn-over at $f = 1/(11\,{\rm y})$ due to binary environmental coupling.
}
\label{fig:timetodet2}
\end{figure}

\subsection{The search for new pulsars}\label{sec:newpsrs}
% Scaling
Finding many pulsars that can be timed to low $\sn$ is another important part of increasing PTA sensitivity.
Only a few low-$\sn$ pulsars ($\lesssim200$\,ns) contribute significantly to the PTAs in \S\ref{sec:ptas} \citep[\eg\ only 6/20 pulsars contributed to the limit of][]{pptascience1}. Millisecond pulsars (MSPs)---pulsars with rotational frequencies of hundreds of Hz---are the only pulsars stable enough for PTA use, and there are ongoing world-wide efforts to find and time these to determine their suitability for PTAs \citep[\eg][]{mspsearch1,mspsearch3,mspsearch2}. This has turned up around 20-30 new millisecond pulsars \emph{per year} in the past 6 years, although less than around five of these per year have been found suitable for PTA use, with $\sn\lesssim1\,\mu$s \citep{mauraIPTA}. However, there remain many potential MSP goldmines. It is estimated that there are $\sim 10^4$ MSPs in the galaxy compared to the several hundred currently known \citep[][]{msp-predictions}. This wealth of MSPs will become readily observable with future telescopes with wide bandwidth and large collecting area, like the Square Kilometre Array \citep[SKA;][]{SKA} and the Five hundred meter Aperture Spherical Telescope \citep[FAST;][]{FAST}. If intrinsic timing noise---particularly red noise---is unsolvable, adding pulsars to one's array is the best way to improve sensitivity to GWs. 

%Note which telescopes are finding new pulsars and mention the rate of MSP discovery per year, and how many MSPs are added to PTAs per year.

\subsection{PTA optimization}\label{sec:optimization}
%Recent work has sought to answer the basic question: 
Should we spend our time observing a few wisely-chosen pulsars, or should we observe as many as possible that can reach a fixed $\sn$ limit? This debate remains a somewhat contentious topic, as its answer depends keenly on the severity of (unknown) levels of long-term intrinsic timing noise, on the type of target one is interested in, and on whether one is aiming for simply detection, or also extensive GW source characterization \citep{blunt-instrument,burt+11,christy+14}. For instance, any PTA aimed at GW detection should spend intensive time observing their few lowest-$\sn$ pulsars \citep{lee+12}. However, this optimization scheme is risky if intrinsic red noise dominates these pulsars over the long term; sensitivity to low-frequency signals like the GWB may be lost, thus arguing for long-term programs on large numbers of pulsars. Likewise, localization and characterization of a discrete GW source benefits from a large array of pulsars well-distributed on the sky. Some guidance on this topic comes from the prediction that the GWB will be discovered first \citep[with a probability of $\sim$70-90\%;][]{rosado-gwb}. However, both cosmic variance and long-term science should also be considered.
%; in 10-20\,yr, PTAs will almost certainly have reached sufficient sensitivity to detect at \emph{least} one CW. 

\subsection{Time to detection}\label{sec:timetodet}
The question of ``when will we detect GWs'' confronts a vast range of uncertainties which we review here.
%PTA sensitivity scaling laws have been derived for some signal classes \citep[\eg][]{jenet+06}.
The anticipated time to detection depends most prominently on the strength of the expected signals, but also depends heavily on the number of pulsars available with low $\sn$ that are used in an array (\S\ref{sec:newpsrs}), on the total observing span of the experiment, and on the prevalence of intrinsic red noise in pulsars' spin variations, which can dampen PTAs' sensitivity to low-frequency signals. \citet{cutler14} derived a generic S/N scaling for any signal of known shape in the case of white-dominant and red-dominant timing noise (see their Eq.\,2.1 and 2.7, respectively).

\citet{siemens+13} demonstrated that scaling laws differ in the weak, intermediate, and strong GW signal regimes, and used these scalings---coupled with predictions of new pulsar discoveries and varying levels of white and red noise---to predict when the NANOGrav timing array might make its first detection of a power-law GWB from BSMBHs (however, it is not clear that the GWB will be a power-law; see Fig.\,\ref{fig:astrophysics}). The result of that analysis, coupled 
%is reproduced in Fig.\,\ref{fig:timetodet}. Note that the 
most recent GWB limit \citep{pptascience2} of $\ayr<1.0\times10^{-15}$, indicates that for NANOGrav the earliest expected detection of the GWB could be between 2019 and 2023, barring the most pessimistic red noise scenario explored by \citet{siemens+13}. 

\citet{taylor+submitted} explored the impact of BMSBH stalling and environment on the time to detection of a GWB (\S\ref{sec:unknowns}), as shown in Fig.\,\ref{fig:timetodet2}. They demonstrate that the expected detection dates will be accelerated with the full IPTA, which can provide immediate access to longer data sets with several low-$\sn$ pulsars. The time to detection will be pushed further into the future if there is a low-frequency turn-over in the GWB (\S\ref{sec:smbhunknowns}), although the expected delay in detection due to a turn-over might only be up to $\sim$2--3\,years. Importantly, Fig.\,\ref{fig:timetodet2} demonstrates how while timing only a few high-precision pulsars may place strict limits on the GWB, a confident detection can only be achieved by timing a large array of pulsars.

Finally, note that it may also be that a turn-over in the GWB spectrum would make the detection of CWs more likely in the near-term, despite delaying a GWB detection; this possibility has yet to be explored.

%The question of time-to-detection is fraught with uncertainties of all kinds and it is probably better to say less than more. However it is also unwise to be overoptimistic (many of our colleagues would qualify). A point that might be of interest is: A PTA which can place a bound of Amax on the GWB could not detect a GWB of amplitude X*Amax with X depends on the number of good pulsars, but is a large number, like sqrt(20) or more. So as we get closer to detection the bounds we are able to place will flatten out. We will notice this flattening years before detection occurs.

\subsection{Anisotropy of the sky}
Until recently, a simplifying assumption had been made that the GWB is formed by a smooth distribution of sources. Now under consideration is whether the sky could in fact have an anisotropic distribution. In this case, one or several discrete GW sources that are \emph{not} resolvable from the background could induce greater GW power in the background than other areas of sky \citep[\eg][]{vik12}. Such might be expected from the natural large-scale clustering in a hierarchically built Universe \citep[\eg][]{hotspots}. Accordingly, methods to limit, detect, and characterize an anisotropic background are being developed \citep{chiaraanisotropic,tayloranisotropic}.

\section{Advances in Structure Formation}\label{sec:smbhunknowns}
%!!! The Astrophysics of Pulsar Timing Arrays
As seen by their starring role in \S\ref{sec:gwsrc} and Figs.\,\ref{fig:avstime}--\ref{fig:cwskyplot}, BSMBHs are the brightest anticipated GW source in the PTA band. BSMBHs are formed during the merger of two massive galaxies, and these facts together decisively bind PTA detection of GWs to studies of galaxy evolution and structure formation in the Universe.
%---comes from the cosmological population of binary supermassive black holes (BSMBHs).

Generally the most massive, nearby pairs should form the dominant contribution to PTA-detectable signals. That is, binary systems with redshift $z\lesssim2$ and chirp mass $\mc>10^8\,\msun$ \citep{sesana+08}. Chirp mass is defined for a binary with masses $m_{\bullet1}$ and $m_{\bullet2}$ as \mbox{$\mc = (m_{\bullet1}m_{\bullet2})^{3/5}/(m_{\bullet1}+m_{\bullet2})^{1/5}$}.

Direct constraints on BSMBH demographics are far from robust, because only a handful of candidate BSMBH systems have been identified. Accordingly, GW predictions are based on theoretical and observational constraints on the evolution of SMBH host galaxies.

As a result of this, GW predictions show a broad range in strength and frequency distribution. In this section we paint a picture of how galaxy evolution is thus decisively a PTA science.

\subsection{Building gravitational waves}
The GWB from BSMBHs is simply the square strain integrated over all discrete BSMBH systems:
\begin{multline}
\label{eq:strain}
h_{c}^{2}(f_r) = \int \int \int \frac{dN_\bullet}{dz ~dm_{\bullet1} ~dq_\bullet ~d(\textnormal{ln} \fgearth)}\\
\times~h_{s}(m_{\bullet1},q,z,\fgearth)^{2} ~dz ~dm_{\bullet1} ~dq_\bullet~,
\end{multline}
here posed as the integration over the differential number of BSMBHs $N_\bullet$ per mass, mass ratio $q$, redshift, and logarithmic frequency interval \citep[\eg][]{sesana+08}. The rotation and sky-averaged strain of a discrete source is
\begin{equation}
\label{eq:hrms}
    h_{s} = \sqrt{\frac{32}{5}}~(\pi \fgrest)^{2/3}~\frac{\left(G \mc\right)^{5/3}}{c^{4}D} ~,
\end{equation}
\citep[\eg][]{petersmathews63}
where $D$ is the proper (co-moving) distance to the binary. 

As previously noted, there is no way to directly measure the distribution of BSMBHs in $z$, $m_{\bullet1}$, and $q_\bullet$, however we can measure the properties of SMBH host galaxies.
Thus the \textit{galaxy} distribution, $dn/(dz\,dM\,dq)$, is produced as a proxy. Subsequently, galaxies in that distribution are populated with SMBHs to then form $h_c(f)$. This can be done via the application of any of the empirical relations that tie black hole mass to host galaxy properties ($M_\bullet$ with bulge mass, stellar velocity dispersion, etc., \eg\ \citealt{mbulge,msigma})

In early work, predictions were derived from numerical dark matter simulations and semi-analytic prescriptions to infer galaxy and SMBH evolution properties \citep[][]{jaffebacker,wyitheloeb,enoki+04,sesana+04,sesana+08}.
Recent work, however, has focused on using the \emph{observed} demographics of the $z\lesssim3$ galaxies that should host SMBHs, to directly infer the expected GWB and CW signals from supermassive binaries \citep[see, \eg,][]{sesana13,ravi+15,simonbs}. The number density of galaxy mergers is formed by combining the galaxy mass function at redshift $z$, $\phi(M,z)$, with the galaxy merger rate; this expression can be constructed as 
\begin{equation}
\frac{dn}{dzdMdq} = \Phi(M,z) \frac{\mathcal{F}(z,M,q)}{\tau(z,M,q)}\frac{dt_{\rm r}}{dz}~,
\end{equation}
where $\mathcal{F}$ is the pair fraction of galaxies, and $\tau$ is the typical merger timescale for a galaxy pair (\eg\ a dynamical friction timescale). The factor $dt_{\rm r}/dz$ converts the proper time rate into a redshift rate. The parameters $\phi$ and $\mathcal{F}$ are observables, but $\tau$ is typically estimated from numerical simulations of galaxy mergers \citep{sesana13}.

%NOTE CHANGES IN $\hc(f)$ by e.g. Sesana (a high frequency turn-over due to low source counts) and Sesana, Ellis, Mcwilliams, Sampson (low-frequency turn-overs due to environmental effects---show the $\hc(f)$ double-power-law.

\begin{figure*}
\centering
\includegraphics[trim=2.5cm 0.75cm 1.25cm 1.25cm,clip,width=0.55\textwidth]{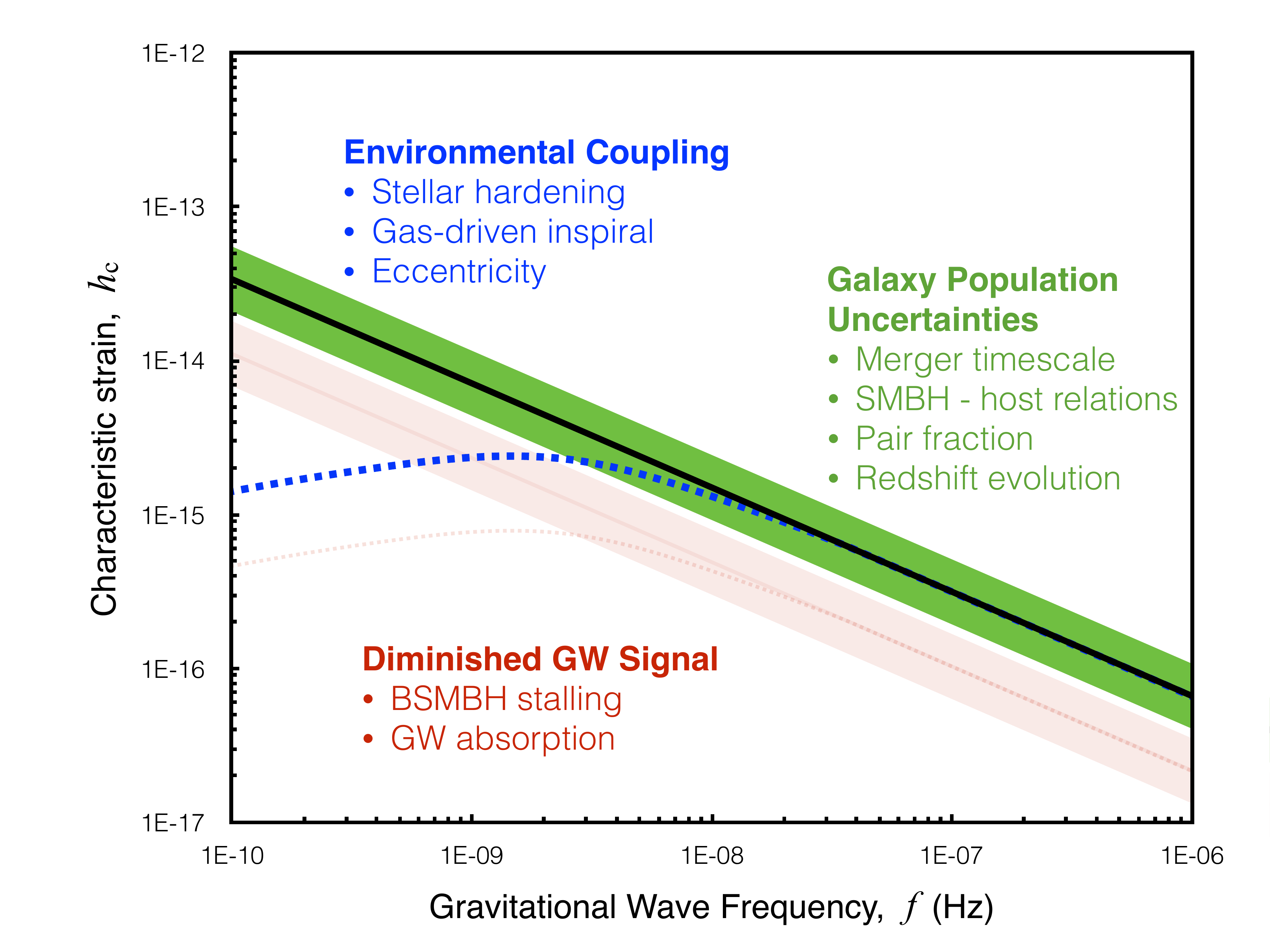}
\caption{A conceptual view of how various uncertainties in the BSMBH population and the GWs we can detect from them can influence the amplitude and shape of the GWB. The line gives the mean of the power-law spectrum of \citet{sesana13}, while the range shows the 68\% range of power-law amplitude predictions based on the variance in observational limits on SMBH host galaxy properties. See \S\ref{sec:unknowns} for a discussion of the effects shown in this figure.}
\label{fig:astrophysics}
\end{figure*}

\subsection{The effect of unknowns}\label{sec:unknowns}
The previous section outlined how the parameters of galaxy evolution relate to PTA measurements of GWs. An added layer of complexity for predicting GW signals from the BSMBH population is a binary's evolutionary path after its host galaxies merge. The basic process is that the SMBHs are first drawn to the merger's center via dynamical friction, form a wide binary and undergo three-body interactions with stars, after which other processes dissipate enough angular momentum for the binary to efficiently evolve to coalescence via gravitational radiation \citep{begelman+80}. If no other processes dissipate the system's energy in the second to last stage, the binary may ``stall,'' enduring a gigayears-long wait before entering the PTA band. On the other hand, three-body interactions and other effects such as gas inflow \citep{mayer} may be so efficient that the binary is not purely driven by GWs through the nHz-$\mu$Hz band, hence changing the expected form of the GW signal. Figures \ref{fig:turnovers} and \ref{fig:astrophysics} show how such ``super-efficient evolution'' could make the expected GWB deviate from a simple power-law at low frequency \citep[see also][]{enoki+07,kocsissesana11,sesana-insights, mcwilliams+14,ravi+14}.

Recent efforts have broken down how uncertainties in galaxy and BSMBH evolution effect GW signals \citep{ravi+15,simonbs,pptascience2}. A conceptual view of the influence of various uncertainties is shown in Figure \ref{fig:astrophysics}. 
%Some studies have explored what the dominant factors of theoretical and observational uncertainty are in GWB predictions.
The below factors are noted approximately in the order of what have the most to least amount of influence on GW signals (strength, rate of occurrance, and form) in the PTA band:
\begin{itemize}
\item \textbf{Effect of environment on BSMBH evolution}: As described above, we have basically no observational data on how BSMBHs couple with their environment during their evolution on scales of $\lesssim$10\,pc. This is the most significant wildcard in GW predictions, although it is only expected to affect the lowest frequencies, $\fgearth\lesssim10^{-8}$\,Hz.
\item \textbf{BSMBH stalling}, as a counter to the previous point, is essentially caused by the non-interaction of the BSMBH with its environment, leading to an exceptionally long time spent at separations $\sim0.1$--$10$\,pc. Long stalling times can potentially cause a large drop in the overall amplitude of the GWB, directly lowering the total number of BSMBHs contributing to the  $dN_\bullet/d(...)$ term in Eq.\,\ref{eq:strain}.
\item \textbf{Eccentricity} redistributes power to higher harmonics in GW frequency and changes the waveform of a CW. If eccentricity is common it would increase the complexity of CW searches, although may raise the probability of CW detection over that of GWB detection \citep{taylor+ecc15,huerta+15}. It may also raise the expected rate of bursts, although that has not yet been investigated. Unless all BSMBH systems are driven by environments to extreme orbits ($e\gtrsim 0.9$), the GWB strength will be diminished, but not by much \citep{enoki+07,ravi+14,huerta+15}.
\item \textbf{Merger timescale, $\tau$}, was shown by \citet{ravi+15} to be the most significant contributor to variance in $\ayr$ predictions; the range of theoretical estimations for this parameter vary by a large factor \citep[\eg][]{kw08,lotz+11}.
\item \textbf{SMBH-host galaxy relations and their redshift evolution} are used to cast galaxy properties to the BSMBH properties used in GW simulations. Measurement error and intrinsic scatter in these vary the GW predictions, however the most significant issues are a) the deficit of direct SMBH measurements at high redshift that forbid a measurement of the redshift-dependence, and b) the deficit of measurements of the highest-mass SMBHs which would contribute most strongly to PTA-detectable signals.
\item \textbf{GW-diminishing effects}: Finally, it has been noted that there may be several other, more exotic, causes of a diminished GW signal. \citet{pptascience2} suggested that their limit was consistent with ``GW absorption'' to other matter in the Universe; while according to \citet{hawking66} this could in principle occur, our current knowledge of cosmological constants imply that in our expanding Universe this should not be the case, as the mean free path is much greater than the GW wavelength, particularly at the redshifts $z\lesssim3$ from which we expect the majority of contributing GW signals from BSMBH systems. 
If gravity instead follows one of many non-general-relativistic theories, this could in principle change the expected signal, although the effect of these on the GWB has not been investigated.
%\item \textbf{Galaxy stellar mass functions} formed from surveys also have intrinsic measurement error and completeness limits, which contribute variance to GW predictions that is notable, however is likely insignificant compared to the above sources of uncertainty \citep{ravi+15,sbs}.
\end{itemize}
Measurement uncertainties in galaxy stellar mass functions and pair fractions appear to contribute relatively little to the variance in GWB simulations \citep{ravi+15,simonbs}.

%\begin{figure*}
%\centering
%\includegraphics[width=0.4\textwidth]{blah.png}
%\caption{!!! A FEW FIGURES SHOWING NANOGRAV AND OTHER RESULST THAT HIT ON THE BSMBH/STRING IMPLICATIONS OF THE GWB LIMITS!!!}
%\label{fig:astrolimits}
%\end{figure*}

\subsection{What PTAs can address}\label{sec:ptaaddress}

Most fundamentally, PTAs are able to limit or measure GW power as a function of frequency. Studies of galaxy mergers and evolution provide constraints on many, but not all, of the constituents of Eq.\,\ref{eq:strain}. Thus, PTAs can infer limits on the unknowns detailed in the previous section. Here we highlight a few recent results that provide examples of how PTA science is impacting galaxy evolution.\footnote{PTAs can limit BMSBH merger densities as a function of mass without any external assumptions or inputs, however these limits are not strongly constraining \citep{hannah}.}

In the basic assumption of a power-law background (Eq.\,\ref{eq:powerlaw}), \citet{simonbs} mapped PTA limits directly to limits on the slope, intercept, and intrinsic scatter of the $M_\bullet$-$M_{\rm bulge}$ relation. The PTA limits of \citet{pptascience2} were discrepant with some $M_\bullet$-$M_{\rm bulge}$ measurements. Simon \& Burke-Spolaor thus demonstrated that BSMBH inspirals stalling on average timescales of $>0.73\,$Gyrs could solve this discrepancy.

%\begin{figure}
%\centering
%\includegraphics[width=0.75\columnwidth]{blah.png}
%\caption{SS limits overlaid on simulations and candidates .... that is, the one I keep putting in my talks and saying it's from my CQG paper. Need to update that figure... maybe can also put in Joe or Sesana's SS list.}
%\label{fig:mm}
%\end{figure}

Employing Bayesian evidence techniques, \citet{9yrNANOlimit} determined that their limit on BSMBHs favors a low-frequency turn-over over a pure power-law spectrum with an odds ratio of 2.2:1 for the model of \citet{sesana13}. This is a model which aimed to estimate the full range of uncertainties in our knowledge of galaxy evolution. That is, if other GW-diminishing effects are not influential (Fig.\,\ref{fig:astrophysics}), the Arzoumanian et al.\ limit has shown that environmental factors do appear to be influencing the binary at frequencies $\lesssim10^{-8}$\,Hz. This work went on to demonstrate how the turn-over frequency of the spectrum can be mapped to the mass density of stars in the galactic core, the accretion rate of the primary black hole from a circumbinary disk, and/or the eccentricity of a BSMBH at binary formation. 

Similarly, \citet{pptascience2} interpreted their limit as favoring either stalled BSMBHs or strong environmental interactions, as their upper limit---the most stringent to date, of $\ayr<1.0\times10^{-15}$---rules out the majority of simplistic power-law models that take into account neither stalling nor environmental coupling.
%limits on the merger rate, and proven that a turn-over is more likely than a power-law spectrum, implying that the environment is in fact influential (recent PPTA,NANO papers---check conclusions in EPTA paper!)
%noted that the host galaxies of low-redshift, high-mass systems relevant to PTAs are reasonably well-constrained by observation, and thus the BSMBH GWB is best predicted from observations of these galaxies \citep[\eg][]{sesanasystematic,mcwilliams14,rosado,ravi+15,simonbs}. Improvements in our knowledge of the source population 

%\begin{itemize}
%\item[a.] One of the biggest issues right now.\\
%   --$>$ Whereas LIGO has specific predictions, summarise here the various efforts to determine how unconstrained variables in cosmology and astrophysics (mass function scalings, gas inflow, m-sigma relations, eccentricity, etc.)
% \item[b.] Astrophysical turn-overs
% \item[c.] The converse: not ``just one number'' MOVE THIS? ``WHAT DO PTAS LIMIT''?\\
% --$>$ Note that it's not just one (or even two) numbers that PTAs will measure (typically seen as GWB strain and amplitude): tie this in to the previous section about how PTA limits (or a detection) can define what we know about cosmological evolution.
%\end{itemize}

\section{Multi-messenger detection}\label{sec:mm}
``Multi-messenger detection'' refers to the detection of an object in both electromagnetic and non-electromagnetic emission. This is a particularly attractive venture as it would open up novel studies of extreme-gravity systems, providing access to physical and astrophysical studies beyond those which could be done with GW or electromagnetic emission detection alone. In addition, like the galaxy evolution studies described in the previous section, this is another symbiotic PTA science; an electromagnetic detection of a BSMBH's periodicity within the nHz-$\mu$Hz band could raise the signal-to-noise ratio of a CW by $\geq$30\% \citep{ellisAAS}, potentially enabling a confident GW detection of that binary. On the other hand, a CW detection reported by PTAs would provide surveyers a targeted swath of sky in which to perform a deep search for a system of relatively well-constrained period, distance, and mass \citep{ellis13bayesBG}.

%(here GWs, although the term can also refer to neutrinos). 
Considering a BSMBH, multi-messenger emission with PTAs would need an electromagnetic marker of a coalescence event for memory, or of a weeks- to decades-period BSMBH for CW detection.  \citet{burkespolaor13} and \citet{schnittman} provide recent reviews of the gamut of electromagnetic emissions that may accompany PTA targets.

There are extensive ongoing efforts to identify BSMBHs at nearly all electromagnetic wavebands \citep[see \eg][for only a few of many examples in the past two decades]{rodriguez+06,eracleous,radiocensus,CRTS}. Some candidates have been identified that lie in the nHz-$\mu$Hz band \citep[\eg][]{OJ287,graham+15,panSTARRS}, however none since 3C66B have been expected to be detectable by PTAs. %(Fig.\,\ref{fig:mm}).
%!!! ADD IN A SS SOURCE FIGURE IF I WANT AFTER REFEREE REPORT !!!

%OUTLINE:
%\begin{itemize}
%\item Define the term multi-messenger (``signature'' vs ``counterpart'').
%\item Of the discrete sources (burst, memory, CW), we need electromagnetic emission that can mark the existence of the binary (i.e.: marking the coalescence for memory, or marking a binary of period few weeks to decades).
%\item \citet{burkespolaor13} and \citet{schnittman} have provided some good reviews of this; see there.
%\item For memory, typical timescales are years at least. So our only chance is the pulsar term, and we have only a small window of time in which the binary must have coalesced for us to even be sensitive to it ($T$ times $N_{\rm pulsars}$).
%\item Many binary candidates are being observed now, some of which are in principle accessible by PTAs.
%\item Simulations tell us that although we expect the GWB first, knowing the binary period can raise the signal-to-noise ratio of a CW by $\geq$30\% \citep{EllisBS or cite Ellis AAS proceeding}.
%\end{itemize}
%(include section if the relevant publications are out by the time of submission)}
%
%\noindent --$>$ How electromagnetic detection might help PTA detection of their GWs.\\
% --$>$ Do some basic analyses for prospects of burst, CW, and memory sources.

%\section{The future} MAKE THIS PART OF CONCLUSIONS?
%\subsection{future PTAs}
%including new telescopes and predictions of new pulsars
%
%Say earlier that PTAs are largely S/N-dominated. Here, explain that things like SKA and FAST will beat down the noise.

\section{Conclusions}
We have summarized the current status of gravitational wave detection with pulsar timing array experiments, and hope to have impressed the extremely broad range in science and expertise that this field involves.

We would like to close with a few concluding thoughts from this review:

%\begin{enumerate}
%\item 
\textbf{Pulsar timing forms a competitive and complementary method of gravitational wave detection.} The earliest expected detection could arise in just a few years (\S\ref{sec:timetodet}), and PTAs' target gravitational waveband and source populations are complementary to those of space-based and ground-based GW experiments (Fig.~\ref{fig:overview}).

%\item 
%\textbf{PTAs are already impacting our knowledge of galaxy evolution} through ...

%\item 
\textbf{Supermassive black hole binaries are likely to be PTAs' dominant target signal}, exceeding the waves expected from other sources like cosmic strings and inflationary GWs. This includes sensitivity to bursting and continuous GW sources, as well as memory events and an ensemble background (\S\ref{sec:gwsrc}).

\textbf{PTAs science is Galaxy Evolution and Binary Supermassive Black Hole science}. As PTAs are cutting into the predicted GWB from BSMBHs, we are beginning to understand how to map GW upper limits directly to limits on the properties of the BSMBH population and their host galaxies (\S\ref{sec:smbhunknowns}). While PTAs can not currently decouple all the parameters, further observational constraints on the galaxy population---as well as the discovery and study of BSMBHs in the gravitational wave regime (\S\ref{sec:mm}), which will reveal their eccentricities and how efficiently these systems evolve---will lead to tight characterization of the BSMBH population.
%. There are significant unknowns in the anticipated strength of all PTA targets, including BSMBH systems 
%PTAs' upper limits on GWs have become a unique and powerful method to limit the underlying physics and astrophysics of these phenomena (\S\ref{sec:gwb},\ref{sec:ptaaddress})
%\end{enumerate}

%\item 
\textbf{PTA sensitivity to GWs is impacted by radio telescope sensitivity, the limited number of known good timing pulsars, and timing noise.} Future telescopes like FAST and the SKA will greatly improve our sensitivity by providing high S/N detections of good timers, as well as probing deeper into our Galaxy to find a large number of PTA-contributing pulsars. The increasing-at-low-frequency red noise exhibited by some timing pulsars (\S\ref{sec:noise}) can cause significant sensitivity issues for long-term ($>10$\,yr) experiments, however by adding more pulsars to an array, red noise might not have a strong impact on our time to detection (\S\ref{sec:timetodet}).

%\item 
\textbf{PTAs are much more than great astrophysics.} Although we did not review non-GW science in this review, pulsar timing contributes fundamentally to other areas of study that we encourage the reader to explore. We briefly mentioned fundamental gravity tests in \S\ref{sec:gwb}; PTAs have also been used to measure terrestrial time standards \citep{hobbs+12} and planetary masses \citep{champion+10}. X-ray timing has also been proposed for use in autonymous spacecraft navigation systems \citep{navigation}.

In conclusion, in the coming years the ongoing timing of pulsars and contributions from new telescopes will rapidly drive PTA science from ``beginning to impact our knowledge of the Universe'' to ``measuring the structure and evolution of the Universe.''

\acknowledgments
The {\sc tempo2} pulsar timing software package was used to generate and analyse data for this paper \citep{tempo2}.
We thank Stefan Oslowski for directing us to single-pulse data from PSR\,J0437--4715.
%, and thank XXX for providing comments on the draft. 
The National Radio Astronomy Observatory is a facility of the National Science Foundation operated under cooperative agreement by Associated Universities, Inc.

%% To help institutions obtain information on the effectiveness of their
%% telescopes, the AAS Journals has created a group of keywords for telescope
%% facilities. A common set of keywords will make these types of searches
%% significantly easier and more accurate. In addition, they will also be
%% useful in linking papers together which utilize the same telescopes
%% within the framework of the National Virtual Observatory.
%% See the AASTeX Web site at http://aastex.aas.org/
%% for information on obtaining the facility keywords.

%% After the acknowledgments section, use the following syntax and the
%% \facility{} macro to list the keywords of facilities used in the research
%% for the paper.  Each keyword will be checked against the master list during
%% copy editing.  Individual instruments or configurations can be provided 
%% in parentheses, after the keyword, but they will not be verified.

%% Appendix material should be preceded with a single \appendix command.
%% There should be a \section command for each appendix. Mark appendix
%% subsections with the same markup you use in the main body of the paper.

%% Each Appendix (indicated with \section) will be lettered A, B, C, etc.
%% The equation counter will reset when it encounters the \appendix
%% command and will number appendix equations (A1), (A2), etc.

%\appendix
%
%\section{Appendix material}
%

\bibliographystyle{mn2e}
\bibliography{biblio}

\clearpage

\end{document}